\documentclass[11pt, a4paper]{article}

\usepackage{jheparxiv,integrable,mathtools,mathrsfs,tikz-cd,amsmath}

\title{Twistors, the ASD Yang-Mills equations and 4d Chern-Simons theory}

\author{Roland Bittleston and David Skinner}

\affiliation{Department of Applied Mathematics \& Theoretical Physics \\
	University of Cambridge \\
	Wilberforce Road \\
	Cambridge CB3 0WA, United Kingdom}

\emailAdd{[r.bittleston, d.b.skinner]@damtp.cam.ac.uk}

\abstract{We show that the approaches to integrable systems via 4d Chern-Simons theory and via symmetry reductions of the anti-self-dual Yang-Mills equations are closely related, at least classically. Following a suggestion of Kevin Costello, we start from holomorphic Chern-Simons theory on twistor space, defined with the help of a meromorphic (3,0)-form $\Omega$. If $\Omega$ is nowhere vanishing, it descends to a theory on 4d space-time with classical equations of motion equivalent to the anti-self-dual Yang-Mills equations. Examples include a 4d analogue of the Wess-Zumino-Witten model and a theory of a Lie algebra valued scalar with a cubic two derivative interaction. Under symmetry reduction, these yield actions for 2d integrable systems. On the other hand, performing the symmetry reduction directly on twistor space reduces holomorphic Chern-Simons theory to the 4d Chern-Simons theory with disorder defects studied by Costello \& Yamazaki. Finally we show that a similar reduction by a single translation leads to a 5d partially holomorphic Chern-Simons theory describing the Bogomolny equations.}

\begin{document}

\maketitle

\section{Introduction} \label{sec:intro}

It has long been known that many integrable systems arise as symmetry reductions of the anti-self-dual Yang-Mills (ASDYM) equations, indeed the ASDYM equations have been proposed as an organising principle for integrable systems. A thorough exposition can be found in \cite{mason1996integrability}. From this perspective, the spectral parameter of an integrable system can naturally be interpreted geometrically as a coordinate on the twistor bundle over space-time. In this paper will be particularly interested in the principal chiral model and Bogomolny equations, which arise as symmetry reductions by non-null translations.\\

Recently an alternative perspective on 2d integrable field theory has garnered much attention. In \cite{costello2019gauge} it was demonstrated that inserting particular classes of order and disorder defects into the theory now known as 4d Chern-Simons (\CSt{4}) theory gives rise to 2d classical integrable field theories. Since then much work has been done enlarging the class of integrable theories which can be obtained in this way. See, {\it e.g.} \cite{delduc2020unifying,costello2020chern,ashwinkumar20204d}. \CSt{4} also treats the spectral parameter as part of the geometry, with the striking consequence that, at the quantum level, anomalies and RG flows can be interpreted geometrically~\cite{costello2017gauge,gaiotto2020integrable}.\\

In this paper we will show that these two perspectives are in fact linked. To achieve this we will exploit a construction proposed by Costello in a very interesting seminar \cite{costello2020topological}. The Penrose-Ward transform identifies anti-self-dual (ASD) connections on Euclidean space with certain holomorphic vector bundles over its twistor space \cite{penrose1984spinors,ward1990twistor}, so it is natural to expect that holomorphic Chern-Simons (HCS) theory on twistor space should be equivalent to ASDYM theory on Euclidean space-time. In writing down a gauge invariant action for HCS one must make a choice of global holomorphic (3,0)-form, but there are none on twistor space. In the context of twistor strings~\cite{witten2004perturbative} this was overcome by instead working on $\cN=4$ super-twistor space ${\CP}^{3|4}$, which is a Calabi-Yau supermanifold. Truncating the superfield expansions gives holomorphic BF-theory on twistor space, a rather more trivial theory than HCS. Costello's insight was that one can work with HCS directly on the non-supersymmetric space by allowing $\Omega$ to be meromorphic instead of holomorphic. This comes at a cost: there is a potential failure of gauge invariance at the location of the poles of $\Omega$, and this must be avoided by imposing appropriate boundary conditions on the fields.\\

Costello claims that for the simplest possible choice of measure, HCS is equivalent to a 4d Wess-Zumino-Witten model (\WZWt{4}) on Euclidean space-time $\E^4$. The classical equation of motion of this theory is Yang's equation - a particular formulation of the ASDYM equations. We will find that different choices of measure lead to a range of actions on space-time many of which, but not all, have equations of motion equivalent to the ASDYM equations. We will refer to 4d theories obtained in this way as integrable, since it is manifest from the twistor description that they can be understood in terms of holomorphic bundles over twistor space. All of the actions we obtain break Lorentz invariance.\\

We will also understand the link between symmetry reduction and \CSt{4}. Quotienting $\E^4$ by a 2-dimensional group of translations turns our 4d integrable theory into a 2d integrable theory. On the other hand, lifting the generator of this symmetry to twistor space, and quotienting there gives \CSt{4}. The process of quotienting introduces disorder defects into \CSt{4}, and the corresponding 2d integrable theory will be the same as that obtained by directly quotienting on space-time. This framework is illustrated in figure \ref{fig:cdintro}.

\begin{figure}[h!] \label{fig:cdintro}
	\centering
	\begin{tikzcd}
	& \begin{tabular}{@{}c@{}}Holomorphic Chern-Simons \\ theory on twistor space \end{tabular} \arrow[dl,"\text{symmetry reduction}",labels=above left] \arrow[dr,"\text{solving along fibres}"] & \\
	\begin{tabular}{@{}c@{}}4d Chern-Simons \\ theory on $\E^2\times{\CP}^1$ \arrow[dr,"\text{solving along fibres}",labels=below left] \end{tabular}  
	& & \begin{tabular}{@{}c@{}} 4d integrable \\ theory on $\E^4$ \end{tabular} \arrow[dl,"\text{symmetry reduction}",labels=below right] \\ & \begin{tabular}{@{}c@{}} 2d integrable \\ theory on $\E^2$ \end{tabular} & 
	\end{tikzcd}
	\caption{A guide to the relationship between integrable systems in 2 and 4 dimensions and Chern-Simons type theories.}
\end{figure}
This paper is organised as follows. In section \ref{sec:sec2} we review the necessary background from Costello's talk, and work through the steps appearing in the above diagram explicitly in the simplest possible example.\\

In section \ref{sec:sec3} we show that by changing the meromorphic $(3,0)$-form on twistor space we can obtain a range of inequivalent actions on $\E^4$ for the ASDYM equations. These include an action proposed by Leznov and Mukhtarov, and Parkes \cite{leznov1987equivalence,leznov1987deformation,parkes1992cubic} and an unpublished action of Mason and Sparling appearing in \cite{mason1996integrability}. We also show that by choosing a meromorphic (3,0)-form on twistor space with zeros we obtain 4d theories which are not equivalent to the ASDYM equations, but nevertheless yield 2d integrable theories under symmetry reduction.\\

In section \ref{sec:ReSt} we discuss the extension to Lorentzian and ultrahyperbolic signatures, and also describe how to obtain standard reality conditions for the resulting integrable theories. Notably, to obtain $\sigma$-models with values in a real from of the gauge group we are obliged to work in ultrahyperbolic signature.\\

Finally in section \ref{sec:HCStoCS5} we consider reductions by 1-dimensional groups of translations. This allows us to obtain a mixed topological-holomorphic 5d Chern-Simons (\CSt{5}) theory describing the Bogomolny equations on $\E^3$. Performing the same reduction in ultrahyperbolic signature gives the 1+2 dimensional chiral model first studied by Ward~\cite{ward1988soliton}.


\section{From holomorphic to 4d Chern-Simons theory} \label{sec:sec2}

In this section we will explicitly work through the steps involved in Figure \ref{fig:cdintro} in the simplest possible case. We begin at the top of the figure, introducing the action for HCS on twistor space $\PT$ proposed by Costello in \cite{costello2020topological}. Costello claims, and we verify, that this theory is equivalent to \WZWt{4} on $\E^4$. We show further that performing a symmetry reduction by a 2d group of translations on $\E^4$ gives the principal chiral model with Wess-Zumino-Witten term (PCM). On the other hand, lifting this group of translations to $\PT$ and performing the reduction there gives \CSt{4} on the quotient. The disorder defects introduced in \CSt{4} are precisely those which give rise to the PCM.


\subsection{Euclidean twistors} \label{subsec:TwiE}

We begin by reviewing twistors in Euclidean signature as described in \cite{atiyah1977instantons,atiyah1978self}, which built on the original works of Penrose \cite{penrose1967twistor,penrose1968twistor,penrose1969solutions}. Throughout this paper we make regular use of spinor index notation and homogeneous coordinates on projective spaces. Further details on our notation and an introduction to homogeneous coordinates can be found in appendices \ref{app:Not} and \ref{app:homcoord} respectively.\\

The twistor space, $\PT$, of complexified space-time, $\C\M^4$, is the total space of the holomorphic vector bundle
\be\cO(1)\oplus\cO(1)\to\CP^1\,. \ee	
It may be provided with homogeneous coordinates $Z^\alpha = (\omega^A,\pi_{A'})$ defined with respect the equivalence relation $Z^\alpha\sim tZ^\alpha$ for $t\in{\mathbb C}^*$. Here $\omega^A = (\omega^0,\omega^1)$ and $\pi_{A'} = (\pi_{0'},\pi_{1'})$ are coordinates on the fibre and base respectively. Note that for $\pi_{A'}$ to determine a point in $\CP^1$ it must be non-vanishing. $Z^\alpha$ can naturally be viewed as an element of $\CP^3$, allowing us to identify
\be \PT = \CP^3\setminus\CP^1\,, \ee
where the $\CP^1$ defined by $\pi_{A'}=0$ has been removed. Points $x^{AA'}\in\C\M^4$ are in bijection with holomorphic lines
\[ \iota_x:\CP^1_x\xhookrightarrow{}\PT,\qquad
\pi_{A'}\mapsto(\omega^A,\pi_{A'}) = (x^{AB'}\pi_{B'},\pi_{A'})\,. \]
As is standard in the twistor literature, here we are implicitly using the Van der Waerden symbols to index space-time coordinates with spinor indices. The action of $\PSL_4(\C)$ on $\CP^3$ induces an action on holomorphic lines which coincides with the action of the complexified conformal group $\SO_6(\C)/\Z_2$ on an appropriate conformal compactification of $\C\M^4$. Those transformations which fix the removed $\CP^1$, and so map $\PT$ to itself, correspond to conformal transformations which descend to $\C\M^4$. In particular, under rotations $\omega^A$ and $\pi_{A'}$ transform as right- and left-handed Weyl spinors respectively.\\

In the first half of this paper we will be concerned with 4-dimensional Euclidean space-time $\E^4$. Complex conjugation in Euclidean signature preserves the handedness of spinors, acting by
\[ \omega^A\mapsto {\hat\omega}^A = (-\overline{\omega^1},\overline{\omega^0})\,,\qquad \pi^{A'}\mapsto {\hat\pi}^{A'} = (-\overline{\pi^{1'}},\overline{\pi^{0'}})\,. \]
These operations obey $\hat{\hat\omega}^A = -\omega^A$ and $\hat{\hat\pi}_{A'} = -\pi_{A'}$ and have no non-vanishing fixed points. They induce the $\SU_2$ invariant inner products
\[ \|\omega\|^2 = [\omega\,\hat\omega] = \varepsilon_{AB}\omega^A\hat\omega^B\,,\qquad \|\pi\|^2 = \la\pi\,\hat\pi\ra = \varepsilon_{A'B'}\pi^{A'}{\hat\pi}^{B'}\,. \]
We can extend complex conjugation to act on $\PT$ in the natural way
\[ Z^\alpha = (\omega^A,\pi_{A'})\mapsto {\hat Z}^\alpha = ({\hat\omega}^A,{\hat\pi}_{A'})\,, \]
and this also has no fixed points. This conjugation provides $\PT$ with a non-holomorphic fibration over $\E^4$, given by
\[ \Pi: \PT\to\E^4\,,\qquad Z^\alpha = (\omega^A,\pi_{A'})\mapsto x^{AA'} = \frac{{\hat\omega}^A\pi^{A'} - \omega^A{\hat\pi}^{A'}}{\|\pi\|^2}\,. \]
Note that points in the image obey $x^{AA'} = \hat x^{AA'}$, and so as claimed lie in the real slice $\E^4\subset\C\M^4$. The fibre over $x\in\E^4$ is $\CP^1_x$. We can think of this fibration as providing a smooth identification between $\PT$ and the left-handed projective spinor bundle over $\E^4$, which we denote by $\PS^+$. Explicitly
\[ \PT\to\PS^+\,,\quad (\omega^A,\pi_{A'})\mapsto (x^{AA'},\pi_{A'})\,. \]
The $\CP^1$ removed from $\CP^3$ to obtain $\PT$ can be viewed as the fibre over the point at infinity in the conformal compactification $S^4$ of $\E^4$. For this reason we will sometimes refer to it as the $\CP^1$ at infinity.\\

We make use of a particularly convenient frame of holomorphic $(0,1)$-forms adapted to the non-holomorphic coordinates $(x^{AA'},\pi_{A'})$ on $\PT$:
\be \bar e^0 = \frac{\la\diff{\hat\pi}\,{\hat\pi}\ra}{\|\pi\|^4} \in\Omega^{0,1}(\PT,\cO(-2))\,,\quad\hat e^A = \frac{\diff x^{AA'}\hat\pi_{A'}}{\|\pi\|^2}\in\Omega^{0,1}(\PT,\cO(-1))\,. \label{eq:formbasis} \ee
These were originally introduced in \cite{boels2007supersymmetric}. Note that they carry holomorphic weight under the scaling $Z^\alpha\mapsto tZ^\alpha$, or equivalently have been twisted by holomorphic line bundles over $\PT$. The dual frame of $(0,1)$-vectors is
\[ {\bar\partial}_0 = \|\pi\|^2\pi^{A'}\partial_{{\hat\pi}^{A'}}\,,\quad {\hat\partial}_A = \pi^{A'}\partial_{x^{AA'}} = \pi^{A'}\p_{AA'} \,. \]
Taking conjugates of the frame of $(0,1)$-forms in equation \eqref{eq:formbasis} gives a basis of $(1,0)$-forms, $\{e^0,e^A\}$. We can similarly conjugate our basis of $(0,1)$-vectors to get the dual frame.\\

There is a canonical holomorphic $3$-form on $\PT$ with holomorphic weight 4 inherited from its natural inclusion into $\CP^3$:
\[ \Diff^3Z = \frac{\varepsilon_{\alpha\beta\gamma\delta}Z^\alpha{\diff Z}^\beta\wedge{\diff Z}^\gamma\wedge{\diff Z}^\delta}{4!} = \frac{\la\diff\pi\,\pi\ra\wedge\diff^2x^{A'B'}\pi_{A'}\pi_{B'}}{2}\in\Omega^{3,0}(\PT,\cO(4))\,. \]


\subsection{Holomorphic Chern-Simons theory on twistor space} \label{subsec:HCS}

The dynamical field of HCS is a partial connection, $\bar\partial + {\bar\cA}$, on a principal $G$-bundle over a 3 complex dimensional manifold $\cW$ with structure group a simple, complex group $G$. The action takes the standard form
\[ S_\HCS[{\bar\cA}] = \frac{1}{2\pi i}\int_\cW\Omega\wedge\mathrm{HCS}({\bar\cA})\,,\]
where
\[ \mathrm{HCS}({\bar\cA}) = \tr\bigg({\bar\cA}\wedge{\bar\partial}{\bar\cA} + \frac{2}{3}{\bar\cA}\wedge{\bar\cA}\wedge{\bar\cA}\bigg) \]
and $\Omega$ is a holomorphic $(3,0)$-form. Here by $\tr$ we mean a $G$-invariant bilinear on $\fg$ proportional to the Killing form.\\

We wish to take $\cW = \PT$, or alternatively its compactification $\CP^3$. To find a suitable $\Omega$, we begin by recalling the canonical holomorphic 3-form on $\CP^3$
\[ \Diff^3Z = \frac{\varepsilon_{\alpha\beta\gamma\delta}Z^\alpha{\diff Z}^\beta\wedge{\diff Z}^\gamma\wedge{\diff Z}^\delta}{4!} = \frac{\la\diff\pi\,\pi\ra\wedge\diff^2x^{A'B'}\pi_{A'}\pi_{B'}}{2}\,. \]
This has holomorphic weight 4, {\it i.e.} takes values in the line bundle $\cO(4)\to\CP^3$, so we cannot simply set $\Omega\overset{!}{=}\Diff^3Z$. To overcome this, Costello chooses a pair of dual twistors $A,B\in(\C^4)^*$  and sets 
\be \label{eqn:firstOmega}
\Omega = \frac{\D^3Z}{(Z\cdot A)^2(Z\cdot B)^2}\,. 
\ee
which is a weightless, but meromorphic $(3,0)$-form. The planes $A\cdot Z=0$ and $B\cdot Z=0$ where $\Omega$ is singular each define a $\CP^2\subset\CP^3$, and they intersect on a $\CP^1$. Removing this $\CP^1$ from $\CP^3$ leaves us with $\PT$. In terms of the coordinates on $\PT$ introduced in subsection \ref{subsec:TwiE} the dual twistors are then given by $A^\alpha = (0^A,\alpha_{A'})$ and $B^\alpha = (0^A,\beta_{A'})$ for some left-handed Weyl spinors $\alpha_{A'}$ and $\beta_{A'}$. We may choose these to be normalized by $\la\alpha\,\beta\ra = \varepsilon_{A'B'}\alpha^{A'}\beta^{B'} = 1$.\\

Allowing poles in $\Omega$ comes at a cost. Varying the action on the support of the classical equations of motion now generates boundary terms at the locations of the poles in $\Omega$. We can eliminate these boundary terms by imposing appropriate conditions on the gauge field.  With $\Omega$ as in~\eqref{eqn:firstOmega} it is sufficient to require that $\bar\cA$ is divisible by $(Z\cdot A)(Z\cdot B)$, that is to say that
\[ 
\bar\cA = (Z\cdot A)(Z\cdot B)\varphi 
\]
for some smooth $\varphi\in \Omega^{0,1}(\PT,\fg\otimes\cO(-2))$. In particular, such an $\bar\cA$ vanishes on each of the planes where $\Omega$ is singular. Since each term in $\mathrm{HCS}(\bar\cA)$ is at least quadratic in $\bar\cA$, the full Lagrangian $\Omega\wedge\mathrm{HCS}(\bar\cA)$ remains regular. For consistency we must also require that this vanishing of $\bar\cA$ on $\{A\cdot Z=0\}\cup\{B\cdot Z=0\}$ is preserved under gauge transformations. Thus, for an  infinitesimal gauge transformation $\bar\cA\mapsto \bar\cA + \bar\p\varepsilon + [\bar\cA,\varepsilon]$ we likewise require that
\[ \varepsilon = (A\cdot Z)(B\cdot Z)\psi \]
for some smooth $\psi\in \Omega^0(\PT,\fg\otimes \cO(-2))$.\\

The classical equations of motion of HCS are
\[ {\bar\cF}(\bar\cA) = {\bar\p}\bar\cA + \bar\cA\wedge\bar\cA = 0 \,. \]
Modulo gauge transformations these are equivalent to the statement that $\bar\p + \bar\cA$ defines a holomorphic $G$-bundle over $\PT$. Using the frame introduced in equation \eqref{eq:formbasis} we may decompose into components along the base and fibres of $\PT = (\cO(1)\oplus\cO(1)\to\CP^1)$:
\[ \bar\cA = \bar\cA_0\bar e^0 + \hat\cA_A\hat e^A\,. \]
In component form the equations of motion read
\begin{align}
{\bar\cF}_{0A}(\bar\cA) &= {\bar\p}_0{\hat\cA}_A - {\hat\p}_A{\bar\cA}_0 + [{\bar\cA}_0,{\hat\cA}_A] = 0\,, \label{eq:HCSeom1}\\ 
{\bar\cF}_{AB}(\bar\cA) &= {\hat\p}_A{\hat\cA}_B - {\hat\p}_B{\hat\cA}_B + [{\hat\cA}_A,{\hat\cA}_B] = 0\,. \label{eq:HCSeom2}
\end{align}
Viewing $\PT$ as a $\CP^1$ bundle over $\E^4$, only the first set of equations involve the component of ${\bar\cA}$ in the direction of the $\CP^1$ fibres, $\bar\cA_0$. We will see that after gauge fixing and imposing only equations~\eqref{eq:HCSeom1}, not~\eqref{eq:HCSeom2}, HCS descends to a local theory on $\E^4$.


\subsection{Effective description as a 4d Wess-Zumino-Witten model} \label{subsec:HCStoWZW4}
To determine the effective space-time action of this theory, and indeed of any CS type theory, we always follow the same basic argument. It was first employed in \cite{costello2019gauge} for \CSt{4}. The first step is to pull back the partial gauge field ${\bar\cA}$ to the fibres of $\PT\xrightarrow{\Pi}\E^4$ via the natural inclusion
\[ \CP_x^1\xhookrightarrow{\iota_x}\PT\xrightarrow{\Pi}\E^4\,, \]
where $\mathrm{Im}(\Pi\circ\iota_x) = \{x\}\subset\E^4$. We write $\bar\cA_x$ for $\iota_x^*\bar\cA = \iota_x^*(\bar\cA_0\bar{e}^0)$. Any complex bundle on $\CP^1$ that is topologically trivial is generically also holomorphically trivial, so 
\[ {\bar\cA}_x = {\sigha}^{-1}{\bar\partial}_{{\CP}_x^1}\sigha \]
for some frame field $\sigha:\PT\to G$. A finite gauge transformation with parameter $g$ acts on the right of $\sigha$ by $\sigha\mapsto\sigha g^{-1}$. Note that the boundary conditions on $g$ prevent us from simply setting $\sigha = {\rm id.}$ everywhere. This is not the only redundancy in $\sigha$, however. We also have the freedom to act on the left by $\sigha\mapsto h\sigha$ provided $\bar\p_{\CP^1_x}h=0$. This condition implies that $h$ is independent of $\pi$, so $h:\E^4\to G$. As we confirm in section \ref{sec:sec3}, $h$ is naturally interpreted as a gauge transformation of the effective description on $\E^4$. Usually we  eliminate both of these redundancies, which amounts to fixing a parametrisation of the moduli space of holomorphic $G$-bundles over $\CP^1_x$ obeying the relevant boundary conditions.\\

In this case we can eliminate the redundancy in $\sigha$ under left translations by fixing $\sigha|_{\pi\sim\beta}=\id$. Then $\sigha|_{\pi\sim\alpha}=\sigma:\E^4\to G$ exhausts the gauge invariant data which can be extracted from $\hat\sigma$. Indeed, $\sigma$ is the holomorphic Wilson line~\cite{mason2010complete,adamo2011scattering} from $\alpha$ to $\beta$
\[ \mathcal{W}_{\alpha\to\beta} = {\rm P}\exp\left(-\frac{1}{2\pi i}\int_{\CP^1_x}\frac{\la\diff\pi\,\pi\ra}{\la\alpha\,\pi\ra\la\pi\,\beta\ra}\wedge{\bar\cA}_x \right) =  \sigha^{-1}|_{\pi\sim\beta}\,\sigha|_{\pi\sim\alpha} = \sigma\,. \]
We can fix the gauge by choosing a specific $\sigha$ which equals $\sigma$ in a neighbourhood of $\alpha$ and is the identity in a neighbourhood of $\beta$.\footnote{In section \ref{sec:sec3} we make an alternative choice of gauge which is equally applicable here. This is achieved by first writing $\bar\cA_x = \la\pi\,\alpha\ra\la\pi\,\beta\ra\varphi_x$ for $\varphi_x$ of holomorphic weight $-2$, and then requiring that $\varphi_x$ be harmonic with respect to the Fubini-Study metric on $\CP_x^1$. Following \cite{woodhouse1985real} we therefore have $\bar\cA_x = \phi\la\pi\,\alpha\ra\la\pi\,\beta\ra\la\diff{\hat\pi}\,\hat\pi\ra/\|\pi\|^4$ for $\phi$ a $\fg$-valued space-time field. It is straightforward to verify that $\sigma = \exp\phi$.} It is also convenient to choose it to be invariant under the natural $\U_1$ action on $\CP^1_x$ preserving $\alpha$ and $\beta$.\\

It is then natural to write
\[ \bar\p + \bar\cA = \sigha^{-1}(\bar\partial + \bar\cA')\sigha \]
for some $\bar\cA'$ obeying $\iota^*_x\bar\cA' = 0$, or in components $\bar\cA' = \hat\cA_A'\hat e^A$. We should think of ${\bar\cA}'$ as being $\bar\cA$ in a `gauge' in which ${\bar\cA}_x$ vanishes.\\

The next step is to solve for ${\bar\cA}'$ in terms of $\sigha$ by imposing the classical equations of motion \eqref{eq:HCSeom2} along the fibres of $\PT\xrightarrow{\Pi}\E^4$, subject to our chosen boundary conditions. We find that
\[ {\bar\cF}_{0A}(\bar\cA) = \sigha^{-1}{\bar\cF}_{0A}(\bar\cA')\sigha = \sigha^{-1}({\bar\p}_0{\hat\cA'}_A)\sigha = 0\,, \]
which imply that the $\hat\cA_A'$ are holomorphic in $\pi$. For the given boundary conditions we find that
\[ \hat\cA_A' = \pi^{A'}A_{AA'} = -\la\pi\,\beta\ra\alpha^{B'}\p_{AB'}\sigma\sigma^{-1}\,. \]
Note that
\[ A_{AA'} = -\beta_{A'}\alpha^{B'}\partial_{AB'}\sigma\,\sigma^{-1} \]
is independent of $\pi$, and coincides with the standard expression for an ASD gauge field in terms of the Yang matrix $\sigma$. (For further details see appendix \ref{app:ASDYM}.)\\

Now that the $\pi$ dependence of $\bar\cA$ is completely determined we can directly perform the integrals over the twistor fibres to reduce the twistor action $S_\HCS[\bar\cA]$ to an effective action on $\E^4$. Similar calculations will be performed in the sequel so we omit the details here. One finds that the resulting 4-dimensional action is
\be S_{\WZW{4}}[\sigma] = \frac{1}{2}\int_{\E^4}\tr(J\wedge{\ast_4J}) + \frac{1}{3}\int_{\E^4\times[0,1]}\mu_{\alpha,\beta}\wedge\tr({\tilde J}^3)\,, \label{eq:WZW4}
\ee
where $J = -\diff\sigma\sigma^{-1}$. Similarly,  ${\tilde J} =  -{\tilde\diff}{\tilde\sigma}{\tilde\sigma}^{-1}$ for $\tilde\sigma$ a smooth homotopy from $\sigma$ to ${\rm id.}$, with $\tilde\diff$  the exterior derivative on $\E^4\times[0,1]$. We have also introduced $\mu_{\alpha,\beta} = \diff^2x^{A'B'}\alpha_{A'}\beta_{B'}$. This action was originally proposed by Donaldson~\cite{donaldson1985anti}, and examined in detail in~\cite{losev1996four}. We refer to it as the 4d Wess-Zumino-Witten model (\WZWt{4}). Varying with respect to $\sigma$ gives
\[ \varepsilon^{AB}\alpha^{A'}\beta^{B'}\partial_{BB'}(\partial_{AA'}\sigma\, \sigma^{-1}) = 0\,,
\]
which we recognise as Yang's equation. (Again, see appendix \ref{app:ASDYM} for further details.) This is consistent with our expectation that HCS on twistor space for this choice of $\Omega$ should have classical equations of motion equivalent to the ASDYM equations on $\E^4$.\\

If we fix $\beta={\hat\alpha}$, then $\mu_{\alpha,\beta} = \mu_{\alpha,\hat\alpha}$ is proportional to the K\"{a}hler form on $\E^4$ in the complex structure determined by $\alpha$. At this point we are not obliged to make this choice of $\beta$, but we will see in section \ref{sec:ReSt} that it will be necessary if we wish to obtain actions for ASD connections taking values in compact real forms of $G$. For the remainder of this subsection we will make this choice.\\

A number of properties of \WZWt{4} are manifest from the twistor perspective:
\begin{itemize}
	\item The choice of spinor $\alpha$ and its conjugate $\hat\alpha$ break the left-handed $\SU_2$ rotation symmetry to $\U_1$, however the right-handed $\SU_2$ is unbroken, leaving an $\U_1\times\SU_2/\Z_2\cong\U_2$ subgroup of the rotation group $\SO_4(\R)$. This is the group of symmetries compatible with a K\"{a}hler structure on $\E^4$. Note that conformal symmetry is broken to the Ponicar\'{e} group by the infinity twistor $I^{\alpha\beta} = \varepsilon^{\alpha\beta\gamma\delta} A_\alpha B_\beta$, so \WZWt{4} is not even classically a CFT.
	\item For this choice of $\Omega$, HCS is invariant under `forbidden' gauge transformations with parameter $g:\PT\to G$ which do not vanish to first order at $\pi\sim\alpha,\hat\alpha$, but instead obey
	\be \alpha^{A'}\p_{AA'}g|_{\pi\sim\alpha} = 0\,, \qquad \hat\alpha^{A'}\p_{AA'}g|_{\pi\sim\hat\alpha} = 0\,. \ee
	We may interpret these conditions as saying that $g|_{\pi\sim\alpha}$ is holomorphic and $g|_{\pi\sim\hat\alpha}$ is antiholomorphic with respect to the complex structure induced on $\E^4$ by $\alpha$. Under this transformation
	\be \sigma\mapsto g|_{\pi\sim\hat\alpha}\sigma g^{-1}|_{\pi\sim\alpha}\,, \ee
	which it straightforward to see is a symmetry of the action \eqref{eq:WZW4}. These are 4d analogues of the familiar loop group symmetries of the 2d Wess-Zumino-Witten model (\WZWt{2}). By taking $g|_{\pi\sim\alpha},g|_{\pi\sim\hat\alpha}$ to be constant we recover the right and left global $G$ symmetries.
	\item Integrating HCS over the orbits of the natural $\U_1$ action on the $\CP^1$ fibres of $\PT$ preserving $\alpha,\hat\alpha$ leads to the K\"{a}hler CS theory of Nair and Schiff~\cite{nair1990kahler,nair1991k,nair1992kahler}. This is similar to the relationship between the \CSt{3} and \CSt{4} realisations of \WZWt{2} as explained in \cite{costello2019gauge}. 
\end{itemize}
It is interesting to note that \WZWt{4} arises as the target space description of the open and heterotic $\cN=2$ string~\cite{ooguri1990self,ooguri1991n,marcus1992n}, which has been conjectured to have connections to twistor theory~\cite{marcus1992harmonic}.\\

This concludes the overlap between the present work and the content of~\cite{costello2020topological}.


\subsection{Symmetry reduction to a principal chiral model with Wess-Zumino-Witten term} \label{subsec:SymRedE}
We now perform a symmetry reduction of $\E^4$ by a 2-dimensional group of translations, $H$. Without loss of generality, we can take the generators of this group to be the complex null vectors
\[ X = \hat\kappa^A\hat\gamma^{A'}\partial_{AA'}\,,\quad \bar X = \kappa^A\gamma^{A'}\partial_{AA'}\,. \]
We will further assume that $\|\kappa\|^2 = \|\gamma\|^2 = 1$ and hence that $\delta(X,\bar X)=1$, where $\delta$ is the flat metric on $\E^4$. Given the dyads $\{\kappa,\hat\kappa\}$ and $\{\gamma,\hat\gamma\}$ we can introduce
\be \label{eq:dbl-null} z = x^{AA'}\kappa_A\gamma_{A'}\,,\quad w = -x^{AA'}\hat\kappa_A\gamma_{A'}\,,\quad \bar w = x^{AA'}\kappa_A\hat\gamma_{A'}\,,\quad \bar z = x^{AA'}\hat\kappa_A\hat\gamma_{A'}\,, \ee
associated double-null complex coordinates on $\E^4$. In terms of these coordinates we have $X=\p_z$ and ${\bar X}=\p_{\bar z}$.\\

In this subsection will apply the symmetry reduction directly to \WZWt{4} \eqref{eq:WZW4}. To achieve this we impose invariance under $X$ and ${\bar X}$ directly on Yang's matrix
\be \cL_X\sigma = \cL_{\bar X}\sigma = 0\label{eq:symredsigma++}\,. \ee
To define what we mean by the Lie derivative we must choose a lift of the action of $H$ to the principal bundle of the gauge theory. In a frame for this $G$-bundle which is invariant under the action of this lift, the Lie derivatives act as they would on a scalar. However, fixing this frame comes at a cost: we lose the ability to perform gauge transformations which do not obey $\cL_X g=\cL_{\bar X}g=0$. This means that the 2d integrable theory we obtain as a symmetry reduction depends on the gauge in which we write down the ASDYM equations, and hence on $\alpha$ and $\hat\alpha$.\\

Under this assumption we learn that $\sigma$ is independent of $z$ and ${\bar z}$. The integral over these coordinates in the action \eqref{eq:WZW4} is then clearly divergent, which can be attributed to the fact that $H$ is non-compact. To overcome this issue we should compactify $\E^4$ and $H\cong\R^2$ to $\E^2\times\T^2$ and $\T^2$, but to avoid the topological subtleties that this would introduce we will simply discard the divergent ${\rm Vol}(H)$ factor. To implement this in practice we can simply contract the bivector $X\wedge{\bar X}$ into the Lagrangian (viewed as a top form) of \WZWt{4}, which has the effect of saturating the form components in the invariant directions. Under the assumed translation invariance \eqref{eq:symredsigma++} the resulting 2-form can be pushed forward to the quotient space. It is convenient to identify the quotient with the transverse section of the orbits $\Sigma = \{z = \bar z= 0\}\subset\E^4.$ We can then replace the pushforward by the quotient map with the pullback by the natural inclusion $\iota_\Sigma:\Sigma\hookrightarrow\E^4$. We find that
\[ (X\wedge{\bar X})\iprod\tr(J\wedge\ast_4J) = \tr(J\wedge\ast_2J)\,,\qquad (X\wedge{\bar X})\iprod\big(\mu_{\alpha,\hat\alpha}\wedge\tr({\tilde J}^3)\big) = k\,\tr(\tilde J^3)
\]
where $k = (X\wedge\bar X) \iprod \mu_{\alpha,\hat\alpha} = \la\alpha\,\gamma\ra\la\hat\alpha\,\hat\gamma\ra + \la\alpha\,\hat\gamma\ra\la\hat\alpha\,\gamma\ra$. Therefore the symmetry reduction of \WZWt{4} by $H$ is the PCM
\be \label{eq:PCM} S_\PCM[\sigma] = \frac{1}{2}\int_{\Sigma}\tr(J\wedge\ast_2J) + \frac{k}{3}\int_{\Sigma\times[0,1]}\tr(\tilde J^3)\,, \ee
where $J$ and ${\tilde J}$ are defined in the usual way. Two reductions are of particular interest:
\begin{enumerate}
	\item If we choose $\alpha\sim\gamma$ or $\alpha\sim\hat\gamma$ then $k=\mp1$ and this action coincides with that of the standard \WZWt{2} model. 
	\item  If $k=0$ we obtain the PCM without WZW term. This occurs when $\alpha$ lies on the equator relative to $\gamma$.
\end{enumerate}
We emphasise that different values of $k$ correspond to different integrable systems. For example, if $\sigma$ is constrained to take values in a compact real form of $\text{SU}_n$, then the PCM without WZW term admits classical solutions extending to $S^2$ \cite{uhlenbeck1989} whereas \WZWt{2} does not. In section \ref{sec:ReSt} we discuss how these realty conditions arise naturally from the gauge theory description.


\subsection{Symmetry reduction to 4d Chern-Simons theory} \label{subsec:HCStoCS4}

While both ASDYM and the PCM are integrable, the spectral parameter is not immediately apparent in their descriptions on $\E^4$ or $\E^2$. This is in contrast to the twistor picture, where the spectral parameter is simply the coordinate $\pi$ on the $\CP^1$ fibres over space-time. In this section we perform the reduction by the translation group $H$ directly on the HCS action on twistor space, keeping the spectral parameter as part of the geometry. The resulting theory is the \CSt{4} description of the PCM given in~\cite{costello2019gauge}.\\

We begin by lifting the action of $H$ to $\PT$, or equivalently by lifting $X$ and $\bar X$ to vector fields $\cX$ and $\bar\cX$ on $\PT$. These are required to obey
\[ \Pi_*\cX = X\,,\quad \Pi_*\bar\cX = \bar X \]
and preserve the complex structure on $\PT$. For translations on $\E^4$ these lifts are completely trivial
\[ \cX = {\hat\kappa}^A{\hat\gamma}^{A'}\p_{AA'} = \p_z\,,\qquad\bar\cX = \kappa^A\gamma^{A'}\p_{AA'} = \p_{\bar z}\,. \]
Viewing $\PT$ as the smooth manifold $\PS^+\cong\E^4\times\CP^1$, we can identify the quotient $H\backslash\PT\cong\E^2\times\CP^1$ with the transverse slice to the orbits $\iota_\cV:\cV=\{z = {\bar z} = 0\}\xhookrightarrow{}\PT$. We abuse notation by continuing to use the coordinates $\{w,\bar w,\pi_{A'}\}$ on $\cV$.\\

To perform the reduction we impose invariance under $\cX$ and $\bar\cX$ directly on the twistor gauge field $\bar\cA$, that is we demand
\be \cL_\cX\bar\cA = \cL_{\bar\cX}\bar\cA = 0\,. \label{eq:symredA++} \ee
Once again, the Lie derivative here is defined by lifting the action of $H$ to the principal $G$ bundle over $\PT$. As in the previous section, we achieve this by working in a frame invariant under the action of $H$, and so the Lie derivatives act as they would on 1-forms. The residual gauge freedom consists of gauge transformations which are invariant under $\cX$ and $\bar\cX$.\\

Under these assumptions the Lagrangian density of HCS is invariant under the action of $H$, and so, as in section \ref{subsec:SymRedE}, the integral over the coordinates $z,{\bar z}$ is divergent. To eliminate this divergence we should compactify in these directions, but instead we will simply discard an overall factor of $\mathrm{Vol}(H)$. In practice this is achieved by contracting the bivector $\cX\wedge\bar\cX$ into the Lagrangian (viewed as a top form) of HCS theory whilst enforcing equation~\eqref{eq:symredA++}. Instead of pushing forward to the quotient we pullback by $\iota_\cV$ and integrate over the representative transverse slice.\\

We find that
\be \begin{aligned}
&\iota_\cV^*\big((\cX\wedge\bar\cX)\iprod(\Omega\wedge\HCS(\bar\cA))\big) = \iota_\cV^*\big((\cX\wedge\bar\cX)\iprod(\Omega\wedge\mathrm{CS}(\bar\cA))\big) \\
&= \frac{\la\pi\,\gamma\ra\la\pi\,\hat\gamma\ra\la\diff\pi\,\pi\ra}{\la\pi\,\alpha\ra^2\la\pi\,\hat\alpha\ra^2}\wedge\mathrm{CS}(A') = \frac{\la\pi\,\gamma\ra\la\pi\,\hat\gamma\ra\la\diff\pi\,\pi\ra}{\la\pi\,\alpha\ra^2\la\pi\,\hat\alpha\ra^2}\wedge\PHCS(A')\,,
\label{eq:SymRedLE} \end{aligned} \ee
where we've defined
\be \begin{aligned}
A' &= \iota^*_\cV\bar\cA - \frac{\la\pi\,\hat\gamma\ra}{\la\pi\,\gamma\ra}\diff w\,\iota^*_\cV(\bar\cX\iprod\bar\cA) + \frac{\la\pi\,\gamma\ra}{\la\pi\,\hat\gamma\ra}\diff\bar w\,\iota^*_\cV(\cX\iprod\bar\cA) \\
&= {\bar e}^0\,\iota^*_\cV\bar\cA_0 + \diff w\,\frac{\iota^*_\cV(\bar\cA^A\kappa_A)}{\la\pi\,\gamma\ra} + \diff\bar w\,\frac{\iota^*_\cV(\bar\cA^A\hat\kappa_A)}{\la\pi\,\hat\gamma\ra}
\label{eq:SymRedAE} \end{aligned} \ee
and
\[ \PHCS(A') = \tr\bigg(A'\diff'A' + \frac{2}{3}A'\wedge A'\wedge A'\bigg) \label{eq:phCSdef} \]
for $\diff' = {\bar\p}_{\CP^1} + \diff_{\E^2} = {\bar e}^0{\bar\p}_0 + \diff w \p_w + \diff{\bar w} \p_{\bar w}$. This $\diff'$ operator can be interpreted in terms of a rather trivial `partially holomorphic structure' on $\cV = \E^2\times\PC^1$. It was essential that in the first line of \eqref{eq:SymRedLE} we replaced $\dbar$ with $\diff$ in the HCS 3-form, since it is not true that $\iota_\cV^*\dbar = \diff'\iota^*_\cV$. We should similarly take care in determining the action of a gauge transformation on the field $A'$. Under an infinitesimal gauge transformation of $\bar\cA$ with parameter $\varepsilon$ obeying $\cL_\cX\varepsilon = \cL_{\bar\cX}\varepsilon = 0$ we have
\[ \begin{aligned}
\delta\bar\cA^A\kappa_A &= \kappa_A{\bar\p}^A\varepsilon + [{\bar\cA}^A\kappa_A,\varepsilon] = \la\pi\,\gamma\ra\p_w\varepsilon + [\cA^A\kappa_A,\varepsilon]\,, \\
\delta\bar\cA^A\hat\kappa_A &= \hat\kappa_A\bar\p^A\varepsilon + [\bar\cA^A\hat\kappa_A,\varepsilon] = \la\pi\,\hat\gamma\ra\p_{\bar w}\varepsilon + [\cA^A\kappa_A,\varepsilon]\,.
\end{aligned} \]
We therefore find that
\[ \delta A' = \diff'\varepsilon + [A',\varepsilon]\,, \]
and so it is natural to interpret $\diff' + A'$ as a `partially holomorphic' connection on a $G$-bundle over $\cV$.\\

From equation \eqref{eq:SymRedAE} we can see that the components $A_{\bar w}'$ and $A_w'$ have simple poles at $\pi\sim\gamma$ and $\hat\gamma$ respectively, and $A'$ is divisible by $\la\pi\,\alpha\ra\la\pi\,\hat\alpha\ra$. Here when we say that $A_w'$ has a simple pole at $\pi\sim\gamma$ we mean that
\[ A_w' = \frac{1}{\la\pi\,\gamma\ra}\varphi_{\bar w}' \]
for $\varphi_{\bar w}'$ smooth and with holomorphic weight 1. It natural to introduce the meromorphic $(1,0)$-form
\[ \omega = \frac{\la\pi\,\gamma\ra\la\pi\,\hat\gamma\ra\la\diff\pi\,\pi\ra}{\la\pi\,\alpha\ra^2\la\pi\,\hat\alpha\ra^2} \]
on $\cV$. We can then write the action on the quotient as
\be \label{eq:CS4} S_{\CS{4}}[A'] = \frac{1}{2\pi i}\int_\cV\omega\wedge\PHCS(A')\,. \ee
In order to make the comparison with \cite{costello2019gauge} as direct as possible we introduce inhomogeneous coordinates on $\CP^1$. In terms of
\[ \zeta = \frac{\la\pi\,\alpha\ra}{\la\pi\,\hat\alpha\ra} \]
we have
\[ \omega = \frac{(\zeta-\zeta_0)(\zeta-\hat\zeta_0)\diff\zeta}{(\zeta_0-\hat\zeta_0)\zeta^2} \]
where $\zeta_0 = \zeta|_{\pi\sim\gamma}$ and $\hat\zeta_0 = \zeta|_{\pi\sim\hat\gamma} = -1/\bar\zeta_0$. The field $A'$ is required to vanish to first order at $\zeta=0,\infty$, though we tolerate simple poles in $A_w'$ and $A_{\bar w}'$ at $\zeta=\zeta_0$ and $\zeta = \hat\zeta_0$ respectively. It is now clear that \eqref{eq:CS4} is the action of \CSt{4} \cite{costello2013supersymmetric}, and in \cite{costello2019gauge} it was argued that for this choice of $\omega$ and corresponding boundary conditions \CSt{4} is equivalent to the PCM on $\Sigma$.\\

That the simple zeros in $\omega$ are located at antipodal points is a consequence of performing the reduction in Euclidean signature. In section \ref{sec:ReSt} we will see how to circumvent this condition, and also discuss compatible reality conditions on $A'$.\\

For completeness we sketch the calculation performed in descending from $\cV$ to $\Sigma$ as presented in \cite{costello2019gauge}. This closely mirrors the computation in subsection \ref{subsec:HCStoWZW4}. First one pulls back $A'$ to the fibres of $\cV\to\Sigma$, and fixes the gauge. The classical equations of motion of \CSt{4} are
\[ F'(A') = \diff'A' + A'\wedge A' = 0\,. \]
Imposing the equation involving the component of $A'$ in the fibre direction subject to our chosen boundary conditions completely fixes the dependence on $\zeta$. We can then directly integrate over $\CP^1$ in \eqref{eq:CS4}. Doing so gives
\[ S_\PCM[\sigma] = \frac{1}{2}\int_{\Sigma^2}\tr(J\wedge\ast_2J) + \frac{k}{3}\int_{\Sigma^2\times[0,1]}\tr(\tilde J^3) \]
for $k=\la\alpha\,\gamma\ra\la\hat\alpha\,\hat\gamma\ra + \la\alpha\,\hat\gamma\ra\la\hat\alpha\,\gamma\ra$. This is the same action that we obtained in subsection \ref{subsec:SymRedE} by performing the symmetry reduction directly on $\E^4$.\\

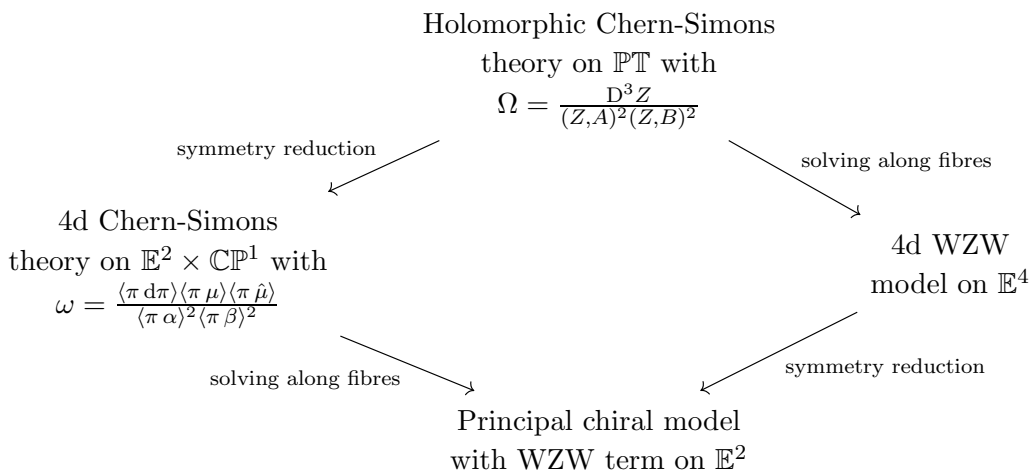
\begin{figure}[h!]
	\centering
	\begin{tikzcd}
	& \begin{tabular}{@{}c@{}} HCS on $\PT$ with \\ $\Omega = \frac{\D^3Z}{(Z\cdot A)^2(Z\cdot B)^2}$ \end{tabular} \arrow[dl,"\text{symmetry reduction}",labels=above left] \arrow[dr,"\text{solving along fibres}"] & \\
	\begin{tabular}{@{}c@{}} \CSt{4} on $\cV$ with \\ $\omega = \frac{(\zeta-\zeta_0)(\zeta-\hat\zeta_0)\diff\zeta}{(\zeta_0-\hat\zeta_0)\zeta^2}$ \arrow[dr,"\text{solving along fibres}",labels=below left] \end{tabular}  
	& & \begin{tabular}{@{}c@{}} \WZWt{4} on $\E^4$ \end{tabular} \arrow[dl,"\text{symmetry reduction}",labels=below right] \\ & \begin{tabular}{@{}c@{}} PCM on $\Sigma$ \end{tabular} & 
	\end{tikzcd}
\caption{A guide to the relationship between the principal chiral model, 4d WZW model and their respective 4d Chern-Simons, twistor progenitors.}
\label{fig:cdWZW4}
\end{figure}

We therefore have the commutative diagram illustrated in figure \ref{fig:cdWZW4}.
\\
The computation performed in this subsection was not sensitive to our choice of $\Omega$. If we quotient HCS theory on $\PT$ with a meromorphic $(3,0)$-form $\Omega = \Phi\D^3Z$ by the 2d group of translations generated by $\cX,\bar\cX$ we will obtain \CSt{4} on the quotient with measure $\omega = \Phi\la\pi\,\gamma\ra\la\pi\,\hat\gamma\ra\la\diff\pi\,\pi\ra$. $A'$ is related to $\bar\cA$ by equation \eqref{eq:SymRedAE}, and so the boundary conditions on $A'$ are inherited from those on $\bar\cA$ with the caveat that we permit poles in $A'_w,A'_{\bar w}$ at $\pi\sim\gamma,\hat\gamma$ respectively. (If there are non-trivial boundary conditions on $\bar\cA$ at $\pi\sim\gamma,\hat\gamma$ then a little more care is needed in determining those induced on $A'$.) This observation will allow us to straightforwardly perform symmetry reductions on $\PT$ in the next section.

 
\section{Twistor actions} \label{sec:sec3}

In this section we will show that HCS on $\PT$ for different choices of $\Omega$ is equivalent to a range of space-time actions on $\E^4$, all describing 4d classically `integrable' field theories.\\

In most of the simple cases we consider here the classical equations of motion for these theories will be equivalent to the ASDYM equations in a particular gauge. We have already seen that one can obtain \WZWt{4}. We will show that it is also possible to recover an action proposed by Leznov and Mukhtarov, and Parkes \cite{leznov1987equivalence,leznov1987deformation,parkes1992cubic,chalmers1996self}, and an unpublished action of Mason and Sparling \cite{mason1996integrability}. In subsection \ref{subsec:CoSi} we consider an example for which the classical equations of motion are not equivalent to the ASDYM equations. Finally in appendix \ref{app:SNovel} we derive novel actions for ASDYM theory, although these explicitly break invariance under both left- and right-handed rotations.


\subsection{Overview} \label{subsec:overview}

We study HCS theory on twistor space
\[ S_\HCS[\bar\cA] = \frac{1}{2\pi i}\int_\PT\Omega\wedge\HCS(\bar\cA)\,, \] 
as introduced in subsection \ref{subsec:HCS}, although for alternative choices of measure
\[ \Omega = \Phi\D^3Z\,. \]
Here $\Phi$ is a meromorphic section of $\cO(-4)$, and is the 4d analogue of the twist function in 2d integrable field theory~\cite{vicedo2019holomorphic}. It generalizes Costello's choice $\Phi = (Z\cdot A)^{-2}(Z\cdot B)^{-2}$ used in subsection~\ref{subsec:HCS}. In this paper we will assume that it factors through $\PT\to\CP^1$, \textit{i.e.}, that $\Phi$ depends only on the $\pi_{A'}$ coordinates on $\PT$. This restriction is motivated by the observation that it will lead to effective space-time actions which do not depend explicitly on the spatial coordinates.\footnote{A  better justification for this restriction is that it ensures invariance under the lifts to $\PT$ of translations and right-handed rotations on $\E^4$. In fact there exist choices of $\Phi$ which depend non-trivially on the $\omega^A$ which are of considerable interest - those which preserve different subgroups of conformal symmetries. We leave investigations of this possibility for future work.} Varying the action on the support of the classical equations of motion will generate `boundary terms' at the locations of the poles in $\Omega$. These must be eliminated by imposing appropriate boundary conditions. Similarly, at zeros of $\Omega$ we can tolerate poles in ${\bar\cA}$ without introducing boundary terms.\\

To obtain the effective space-time description of HCS we will always adopt the same basic approach used by Costello in \cite{costello2020topological} and outlined in section \ref{subsec:HCStoWZW4}. The first step is to fix the gauge on the fibres of $\PT\xrightarrow{\Pi}\E^4$ by writing
\[ \bar\cA_x = \sigha^{-1}{\bar\partial}_{\CP^1_x}\sigha \]
for an appropriate choice of frame field $\sigha:\PT\to G$. $\sigha$ is determined by a map from $\E^4$ to the moduli space of $G$-bundles on $\CP^1_x$ subject to our chosen boundary conditions. In the cases we consider this is always some finite dimensional complex manifold. It is then natural to write
\[ \bar\p + \bar\cA = \sigha^{-1}(\bar\p + \bar\cA')\sigha \]
for some $\bar\cA'$ obeying $\iota^*_x\bar\cA' = 0$, or in components $\bar\cA' = \hat e^A\hat\cA_A'$. We should think of $\bar\cA'$ as being $\bar\cA$ in a `gauge' in which ${\bar\cA}_x$ vanishes. The classical equations of motion involving the component of $\bar\cA$ along the twistor fibres then imply that $\hat\cA_A'$ are meromorphic as sections of $\cO(1)\to\CP^1$. $\bar\cA'$ is then uniquely determined in terms of $\sigha$ by our boundary conditions.\\

At this point the dependence of ${\bar\cA}$ on $\pi$ is completely fixed, and we can integrate over the twistor fibres to get an effective space-time action. Its classical equations of motion will imply
\be\label{eq:effective} 
{\bar\cF}_{AB}(\bar\cA') = 0\,.
\ee
There are two natural ways to interpret this equation.\\

In many cases, equation \eqref{eq:effective} coincides with the ASDYM equations. If we choose $\Phi$ so that it is nowhere vanishing, then $\bar\cA$ will be without poles.\footnote{It is technically possible to allow poles in $\bar\cA$ at points where $\Phi$ is non-vanishing. The residue at such a pole needs to be fixed so that the boundary terms generated when varying the action disappear. In fact under symmetry reductions of HCS by null translations boundary conditions of this form can arise naturally in the resulting \CSt{4}. We will not consider such boundary conditions or reductions in this paper.} Under this assumption we learn that $\hat\cA_A'$ is globally holomorphic in $\pi$, and therefore that
\[ \bar\cA' = {\hat e}^A{\hat\cA}'_A = {\hat e}^A\pi^{A'}A_{AA'}\,, \]
with $A_{AA'}$ independent of $\pi$. In terms of $\nabla_{AA'} = \partial_{AA'} + A_{AA'}$ the equations of motion of the effective space-time theory are
\[ {\bar\cF}_{AB}(\bar\cA') = \pi^{A'}\pi^{B'}[\nabla_{AA'},\nabla_{BB'}] = 0\,, \]
which are the ASDYM equations. This is what we expected from the Penrose-Ward transformation. The boundary conditions on $\bar\cA$ constrain $A$, and these constraints can naturally be interpreted as gauge fixing conditions on the ASD connection $A$. These are in general only attainable for an ASD connection, so parts of the ASDYM equations are actually encoded in the gauge fixing. The equations of motion we obtain from the effective action on $\E^4$ will imply the remainder of the ASDYM equations in this gauge.\\

A more general interpretation of~\eqref{eq:effective} is as the zero-curvature equation of a 4d analogue of the Lax connection of a 2d integrable system. We write
\[ {\hat\scrL}_A = {\hat\cA}_A\,, \]
for the Lax connection, whereupon
\[ {\hat\partial}^A{\hat\scrL}_A + \frac{1}{2}[{\hat\scrL}^A,{\hat\scrL}_A] = 0 \]
is the zero-curvature equation. This interpretation has the advantage of extending to the case where $\Phi$ has zeros. Under symmetry reduction $\hat\scrL_A$ pushes forward to the Lax connection of a 2d integrable system.


\subsection{LMP action} \label{subsec:LMP}

In this section we show that taking $\Omega$ to have a fourth order pole leads to a space-time action for ASDYM theory originally proposed by Leznov and Mukhtarov, and Parkes \cite{leznov1987equivalence,leznov1987deformation,parkes1992cubic}. \\

We choose the meromorphic $(3,0)$-form
\[ \Omega = \frac{\Diff^3Z}{(Z\cdot A)^4} = \frac{\la\diff\pi\,\pi\ra\wedge\diff^2x^{A'B'}\pi_{A'}\pi_{B'}}{2\la\pi\,\alpha\ra^4} \]
where for later convenience we assume that $\|\alpha\|^2=1$. Note that we are implicitly assuming without loss of generality that the locus $Z\cdot A=0$ intersects the $\CP^1$ at infinity. Varying $S_\HCS$ on the support of the classical equations of motion leads to a term
\[ \begin{aligned}
&\delta S_\HCS = \frac{1}{2\pi i}\int_\PT\Omega\wedge\bar\p\tr(\delta\bar\cA\wedge\bar\cA) = \frac{1}{2\pi i}\int_\PT\bar\p\Omega\wedge\tr(\delta\bar\cA\wedge\bar\cA) \\
&= \frac{1}{4\pi i}\int_{\CP^1}\bar\p_{\CP^1}\bigg(\frac{\la\diff\pi\,\pi\ra}{\la\pi\,\alpha\ra^4}\bigg)\int_{\E^4}\diff^2 x^{A'B'}\pi_{A'}\pi_{B'}\wedge\tr(\delta\bar\cA\wedge\bar\cA)\,.
\end{aligned} \]
We can view the integral over $\CP^1$ as picking out the coefficient of $\la\pi\,\alpha\ra^3$ in
\[ \frac{1}{2}\int_{\E^4}\diff^2x^{A'B'}\pi_{A'}\pi_{B'}\wedge\tr(\delta\bar\cA\wedge\bar\cA)\,. \]
By requiring $\bar\cA$ to be divisible by $(Z\cdot A)^2 = \la\pi\,\alpha\ra^2$, {\it i.e.}, to equal
\[ \bar\cA = \la\pi\,\alpha\ra^2\varphi \]
for $\varphi\in\Omega^{0,1}(\PT,\fg\otimes\cO(-2))$ smooth, this term can be eliminated. Infinitesimal gauge transformations by $\varepsilon$ are also taken to be divisible by $\la\pi\,\alpha\ra^2$ in order to preserve this boundary condition.\\

To determine the effective action on space-time we follow the procedure outlined in section \ref{subsec:overview}, although here we employ a different method to fix the gauge than that adopted in \cite{costello2019gauge}. The first step is to pullback $\bar\cA$ via the inclusion $\CP^1_x\xhookrightarrow{\iota_x}\PT$ to get $\bar\cA_x = \iota^*_x\bar\cA$. Our boundary conditions guarantee that $\bar\cA_x = \la\pi\,\alpha\ra^2\varphi_x$ for $\varphi_x$ of holomorphic weight $-2$ and we require that $\varphi_x$ be harmonic with respect to the Fubini-Study metric on $\CP^1_x$. This gives~\cite{woodhouse1985real}
\[ 
\varphi_x = -\frac{\phi\la\diff\hat\pi\,\hat\pi\ra}{\|\pi\|^4} \]
for $\phi:\E^4\to\fg$ depending on $x$ only. Equivalently
\[ \bar\cA_x = -\frac{\phi\la\pi\,\alpha\ra^2\la\diff\hat\pi\,\hat\pi\ra}{\|\pi\|^4}= -\phi\la\pi\,\alpha\ra^2{\bar e}^0\,. \]
Introducing $\sigha$ in the usual way by
\[ {\bar\cA}_x = \sigha^{-1}{\bar\p}_{\CP^1_x}\sigha\,, \]
and fixing the redundancy $\sigha\to h\sigha$ by requiring that $\sigha|_{\pi\sim\alpha} = \id$ for all $x$, we find that  
\be \label{eq:LMPsigha} \sigha = \exp\left(-\frac{\la\pi\,\alpha\ra\la{\hat\pi}\,\alpha\ra\phi}{\|\pi\|^2}\right)\,. \ee
Conversely, the field $\phi$ may be extracted from $\sigha$ as
\[ \phi = -\frac{\la\pi\,\hat\alpha\ra^2\hat\pi^{A'}}{\la\pi\,\hat\pi\ra}\sigha^{-1}\p_{\pi^{A'}}\sigha|_{\pi\sim\alpha} \]
It exhausts the gauge invariant data which can be extracted from ${\bar\cA}_x$. \\

Solving the equations of motion involving $\bar\cA_0$ we find that
\[ \label{eq:FullAHarmonicGauge} \bar\cA = \sigha^{-1}\bar\p\sigha + \sigha^{-1}\bar\cA'\sigha \]
where ${\bar\cA}' = {\hat e}^A\pi^{A'}A_{AA'}$ for $A$ depending only on $x$, since $\Omega$ is nowhere vanishing. The next step is to fix $A$ in terms of $\phi$ using the boundary conditions. We begin by noting that
\begin{equation} \begin{aligned} \label{eq:expandJ}
-\diff\sigha\sigha^{-1} &= \frac{\la\pi\,\alpha\ra\la{\hat\pi}\,\alpha\ra}{\|\pi\|^2}\diff\phi + \frac{\la\pi\,\alpha\ra^2\la\hat\pi\,\alpha\ra^2}{2\|\pi\|^4}[\diff\phi,\phi] + \cO\big(\la\pi\,\alpha\ra^3\big) \\
&= - \frac{\la\pi\,\alpha\ra}{\la\pi\,\hat\alpha\ra}\diff\phi + \frac{\la\pi\,\alpha\ra^2}{2\la\pi\,\hat\alpha\ra^2}[\diff\phi,\phi] + \cO\big(\la\pi\,\alpha\ra^3\big)\,,
\end{aligned} \end{equation}
where $\diff = \diff_{\E^4}$ denotes the exterior derivative on $\E^4$ and $\cO\big(\la\pi\,\alpha\ra^n\big)$ indicates a term which vanishes to order $n$ at $\pi\sim\alpha$. Consider
\begin{align*}
\sigha\hat\cA_A\sigha^{-1} &= \pi^{A'}(\p_{AA'}\sigha\sigha^{-1} + A_{AA'})\,, \\
&= -\frac{\la\pi\,\alpha\ra\la\hat\pi\,\alpha\ra}{\|\pi\|^2}\pi^{A'}\p_{AA'}\phi + \pi^{A'}A_{AA'} + \cO\big(\la\pi\,\alpha\ra^2\big)\,, \\
&= \la\pi\,{\hat\alpha}\ra \alpha^{A'}A_{AA'} + \la\pi\,\alpha\ra(\alpha^{A'}\p_{AA'}\phi - {\hat\alpha}^{A'}A_{AA'}) + \cO\big(\la\pi\,\alpha\ra^2\big)\,.
\end{align*}
For $\bar\cA$ to be divisible by $\la\pi\,\alpha\ra^2$ we require that
\[ \alpha^{A'}A_{AA'} = 0\,,\qquad
{\hat\alpha}^{A'}A_{AA'} = \alpha^{A'}\p_{AA'}\phi \,. \]
The solution to the first is $A_{AA'}\sim\alpha_{A'}$, and then the second implies
\[ A_{AA'} = -\alpha_{A'}\alpha^{B'}\partial_{AB'}\phi\,. \label{eq:LMPA} \]
A generic ASD gauge field can be brought into this form by a gauge transformation, and we refer to it as LMP gauge. (See appendix~\ref{app:ASDYM} for details.)\\

We now show that the twistor HCS action itself reduces to the LMP action for ASDYM theory once we evaluate it on our expression for $\bar\cA$. We first observe that 
\[ \HCS(X+Y) = \HCS(X) + 2\tr({\bar\cF}(X)Y) - {\bar\p}\,\tr(XY) + 2\tr(XY^2) + \HCS(Y)\,. \]
Defining $\bar\cJ = -{\bar\p}\sigha\sigha^{-1}$, and letting $X = \sigha^{-1}{\bar\p}\sigha = -\sigha^{-1}{\bar\cJ}\sigha$ and $Y = \sigha^{-1}{\bar\cA}'\sigha$, we have
\[ \HCS(\bar\cA) = \frac{1}{3}\tr({\bar\cJ}^3) + {\bar\p}\,\tr(\bar\cJ{\bar\cA}') - 2\tr({\bar\cJ}{\bar\cA}'^2) + \tr(\sigha^{-1}{\bar\cA}'\sigha{\bar\p}(\sigha^{-1}{\bar\cA'}\sigha)). \]
For the final term to contribute to the action, $\bar\p$ must act as ${\bar e}^0{\bar\partial}_0$, since ${\bar\cA}'$ has no ${\bar e}^0$ component. Furthermore, recalling that ${\bar\partial_0}{\hat\cA}_{A}'=0$, this ${\bar\p}$-operator must in fact act on the $\sigha$s, so
\[ \tr(\sigha^{-1}{\bar\cA}'\sigha{\bar\p}(\sigha^{-1}{\bar\cA}'\sigha)) = -2\tr({\bar\p}\sigha\sigha^{-1}{\bar\cA}'^2) = 2\tr({\bar\cJ}{\bar\cA}'^2)\,, \]
which cancels the penultimate term. We conclude that
\be \label{eq:CSexpanded} \HCS(\bar\cA) = \frac{1}{3}\tr\big({\bar\cJ}^3) + {\bar\p}\,\tr(\bar\cJ{\bar\cA}')\,. \ee
Let's begin by considering the second of these terms. Its contribution to the action is
\[ \begin{aligned}
&\frac{1}{2\pi i}\int_\PT\Omega\wedge\bar\p\tr(\bar\cJ\wedge\bar\cA') = \frac{1}{2\pi i}\int_\PT\bar\p\Omega\wedge\tr(\bar\cJ\wedge\bar\cA') \\
&= \frac{1}{4\pi i}\int_{\CP^1}\bar\p_{\CP^1}\bigg(\frac{\la\diff\pi\,\pi\ra}{\la\pi\,\alpha\ra^4}\bigg)\int_{\E^4}\diff^2 x^{A'B'}\pi_{A'}\pi_{B'}\wedge\tr(\bar\cJ\wedge\bar\cA')\,.
\end{aligned} \]
We can interpret the integral over $\CP^1$ as extracting the coefficient of $\la\pi\,\alpha\ra^3$ in
\be \label{eq:LMPalg2} \frac{1}{2}\int_{\E^4}\diff^2 x^{A'B'}\pi_{A'}\pi_{B'}\wedge\tr(\bar\cJ\wedge\bar\cA')\,. \ee
To compute it we first note that
\[ \diff^2x^{A'B'}\pi_{A'}\pi_{B'}\wedge\bar e^A\wedge\bar e^B = - 2\varepsilon^{AB}\vol_\delta\,, \]
allowing us to rewrite \eqref{eq:LMPalg2} as
\[
\int_{\E^4}\vol_\delta\,\tr(\hat\cJ^A\hat\cA'_A)\,.
\]
We therefore need to expand $\tr(\hat\cJ^A\hat\cA_A')$ to order $\la\pi\,\alpha\ra^3$. From equation \eqref{eq:LMPA} we know that $\hat\cA'_A$ vanishes to first order at $\pi\sim\alpha$, so it sufficient to expand $\bar\cJ$ to second order in $\la\pi\,\alpha\ra$. From \eqref{eq:expandJ}
\begin{equation} \begin{aligned} \label{eq:LMPalg3} \hat\cJ_A &= -\pi^{A'}\p_{AA'}\sigha\sigha^{-1} \\
&= - \la\pi\,\alpha\ra\alpha^{A'}\p_{AA'}\phi + \frac{\la\pi\,\alpha\ra^2}{\la\pi\,\hat\alpha\ra}\Big(\hat\alpha^{A'}\p_{AA'}\phi + \frac{1}{2}[\alpha^{A'}\p_{AA'}\phi,\phi]\Big)  + \cO\big(\la\pi\,\alpha\ra^3\big)\,. \end{aligned} \end{equation}
The coefficient of $\la\pi\,\alpha\ra^3$ in $\tr(\hat\cJ^A\hat\cA_A')$ is therefore
\begin{align*}
\varepsilon^{AB}\tr\bigg(\Big(\hat\alpha^{A'}\p_{AA'}\phi + \frac{1}{2}[\alpha^{A'}\p_{AA'}\phi,\phi]\Big)\alpha^{B'}\p_{BB'}\phi\bigg)\,.
\end{align*}
We conclude that contribution of \eqref{eq:LMPalg2} to the effective space-time action is
\begin{equation} \begin{aligned} \label{eq:LMPkin} &\frac{1}{2}\int_{\E^4}\vol_\delta\,\varepsilon^{AB}\Big(\varepsilon^{A'B'}\tr(\p_{AA'}\phi\p_{BB'}\phi) - \alpha^{A'}\alpha^{B'}\tr(\phi[\p_{AA'}\phi,\p_{BB'}\phi])\Big) \\ &= \frac{1}{2}\int_{\E^4}\tr\big(\diff\phi\wedge\ast\diff\phi + \mu_{\alpha,\alpha}\wedge\phi\diff\phi\wedge\diff\phi\big)\,,
\end{aligned} \end{equation}
where $\mu_{\alpha,\alpha} = \diff^2x^{A'B'}\alpha_{A'}\alpha_{B'}$. Now let's turn out attention to the first term in equation \eqref{eq:CSexpanded}, given by
\be \label{eq:LMPalg4}
\frac{1}{12\pi i}\int_\PT\Omega\wedge\tr(\bar\cJ^3) = \frac{1}{4\pi i}\int_{\CP^1}\frac{\la\diff\pi\,\pi\ra\wedge{\bar e}^0}{\la\pi\,\alpha\ra^4}\int_{\E^4}\diff^2x^{A'B'}\pi_{A'}\pi_{B'}\,\tr(\bar\cJ_0\bar\cJ^2)\,.
\ee
Consider the $\U_1$ action on $\CP^1$ preserving $\alpha,\hat\alpha$, acting as $\la\pi\,\alpha\ra\mapsto e^{i\theta}\la\pi\,\alpha\ra$, $\la\pi\,\hat\alpha\ra\mapsto\la\pi\,\hat\alpha\ra$. Under this action, the argument of the exponential in the definition of $\sigha$ has positive charge \eqref{eq:LMPsigha}. We deduce that expanding out $\bar\cJ$ using Dunhamel's formula generates terms which also have positive charge. The integral over the orbits of the $\U_1$ action will pick out the invariant part of the integrand. It is therefore clear that it only receives contributions from the coefficient of $\la\pi\,\alpha\ra^4$ in
\[
\frac{1}{2}\int_{\E^4}\diff^2x^{A'B'}\pi_{A'}\pi_{B'}\wedge\tr(\bar\cJ_0\bar\cJ^2)\,.
\]
Now, from \eqref{eq:LMPsigha} $\bar\cJ_0 = \la\pi\,\alpha\ra^2\phi + \cO\big(\la\pi\,\alpha\ra^3\big)$, and from \eqref{eq:LMPalg3} $\hat\cJ_A = -\la\pi\,\alpha\ra\alpha^{A'}\p_{AA'}\phi + \cO\big(\la\pi\,\alpha\ra^2\big)$. We therefore conclude that \eqref{eq:LMPalg4} contributes
\begin{equation} \begin{aligned} \label{eq:LMPint} &-\frac{1}{2\pi i}\int_{\CP^1}\la\diff\pi\,\pi\ra\wedge\bar e^0\int_{\E^4}\vol_\delta\,\varepsilon^{AB}\alpha^{A'}\alpha^{B'}\tr(\phi\p_{AA'}\phi\p_{BB'}\phi) \\ &\qquad = \frac{1}{6}\int_{\E^4}\vol_\delta\,\varepsilon^{AB}\alpha^{A'}\alpha^{B'}\tr(\phi[\p_{AA'}\phi,\p_{BB'}\phi]) = -\frac{1}{6}\int_{\E^4}\mu_{\alpha,\alpha}\wedge\tr(\phi\diff\phi\wedge\diff\phi)\,. \end{aligned} \end{equation}
Combining equations \eqref{eq:LMPkin} \& \eqref{eq:LMPint} gives the LMP action
\be \label{eq:SLMP} S_\mathrm{LMP}[\phi] = \int_{\E^4}\tr\bigg(\frac{1}{2}\diff\phi\wedge\ast\diff\phi + \frac{1}{3}\mu_{\alpha,\alpha}\wedge\phi\diff\phi\wedge\diff\phi\bigg)\,, \ee
originally appearing in \cite{leznov1987equivalence,leznov1987deformation,parkes1992cubic}. Its classical equation of motion is
\[ \diff\ast\diff\phi = \mu_{\alpha,\alpha}\wedge\diff\phi\wedge\diff\phi\,, \]
which implies the ASDYM equations for $A_{AA'} = -\alpha_{A'}\alpha^{B'}\partial_{AB'}\phi$ (see appendix \ref{app:ASDYM} for further details).\\

We make the following observations:
\begin{itemize}
	\item Choosing $\Omega$ to have a fourth order pole clearly requires the fewest arbitrary choices of dual twistors, and so breaks conformal invariance in the least damaging way. Indeed, the Lagrangian of HCS is invariant under the subgroup of the complexified conformal group $\PSL_4(\C)$ preserving the dual twistor $A$. This includes all translations and right handed rotations, but also the combined dilations and left handed rotations acting as
	\be \label{eq:hiddenLMPdilation} \delta x^{AA'} = \frac{1}{2}(\alpha^{A'}\beta_{B'} - 3\beta^{A'}\alpha_{B'})x^{AB'}\,,\qquad \delta\pi_{A'} = \alpha_{A'}\beta^{B'}\pi_{B'}\,, \ee
	and the special conformal transformations acting as
	\be \label{eq:hiddenLMPspecial} \delta x^{AA'} = \frac{1}{2}x^2\lambda^A\alpha^{A'} - x^{AA'}x^{BB'}\lambda_B\alpha_{B'}\,,\qquad \delta\pi_{A'} = \alpha_{A'}x^{BB'}\lambda_B\pi_{B'}\,, \ee
	for left- and right-handed spinors $\beta$ and $\lambda$ respectively. Unfortunately in Euclidean signature the reality conditions on $x$ mean that only the translations and right handed rotations are realised as symmetries of the LMP action. We shall see in section \ref{sec:ReSt} that LMP action only admits natural reality conditions on $\phi$ in ultrahyperbolic signature, for which subgroups of all of the above space-time symmetries survive.
	
	\item In \cite{chalmers1996self} it was observed that a supersymmetric (SUSY) analogue of the LMP action could be used to describe ASD $\cN=4$ super Yang-Mills. This arises from a twistor action
	\be \frac{1}{2\pi i}\int_{\CP^3\times\C^{0|4}}\frac{\D^3Z\wedge\diff^4\chi}{\la\pi\,\alpha\ra^4}\wedge\mathrm{HCS}(\bar\cA) \ee
	where $\bar\cA$ is required to be divisible by $\la\pi\,\alpha\ra^2$ as above. Here $\C^{0|4}$ refers to the complex 4 dimensional odd super vector space with coordinates $\chi^m$ for $m=1,\dots,4$. The SUSY LMP action was demonstrated in \cite{chalmers1996self} to be equivalent to the standard action for ASD $\cN=4$ super Yang-Mills on space-time. At the level of the Lagrangian this equivalence is clear from the twistor description, since we can rewrite the above as an integral over $\CP^{3|4}$ by mapping
	\be \chi^a \mapsto \psi^a = \la\pi\,\alpha\ra\chi^a\,. \ee
	This recovers the familiar twistor action for ASD $\cN=4$ super Yang-Mills of Witten \cite{witten2004perturbative,boels2007supersymmetric}. Less clear is why the boundary conditions on $\bar\cA$ in the two descriptions are equivalent. (A similar supersymmetric analogue of \WZWt{4} was also identified in \cite{chalmers1996self}, which can presumably also be written on twistor space using a minor modification of the construction described in subsection \ref{subsec:HCStoWZW4}.)
\end{itemize}
It is natural to ask what we obtain if we perform a symmetry reduction by the 2d group of translations generated by $X = \hat\kappa^A\hat\gamma^{A'}\p_{AA'}$, $\bar X = \kappa^A\gamma^{A'}\p_{AA'}$. Following the same procedure outlined in section \ref{subsec:SymRedE} this reduction gives
\be \label{eq:SSRCSE2} S_\mathrm{PDPCM}[\phi] = \int_\Sigma\tr\bigg(\frac{1}{2}\diff\phi\wedge\ast_2\diff\phi + \frac{k}{3}\phi\diff\phi\wedge\diff\phi\bigg)\,, \ee
where $k = (X\wedge\bar X)\iprod \mu_{\alpha,\alpha} = 2\la\alpha\,\gamma\ra\la\alpha\,\hat\gamma\ra$. This is the pseudodual of the PCM \cite{curtright1994currents,nappi1980some,zakharov1978relativistically}. For $k\neq0$, or equivalently $\alpha\not\sim\gamma,\hat\gamma$, its equation of motion,
\[ \diff\ast_2\diff\phi = k\phi\wedge\phi\,, \]
is equivalent to the classical equation of motion of the PCM without WZW term. We can also perform this reduction directly on twistor space. Doing so yields \CSt{4} on $\cV = \Sigma\times\CP^1$ with
\[ \omega = \frac{\la\pi\,\gamma\ra\la\pi\,\hat\gamma\ra\la\diff\pi\,\pi\ra}{\la\pi\,\alpha\ra^4} \]
and the obvious boundary conditions on $A'$. \CSt{4} for this choice of $\omega$ has not been studied in the literature, but it is straightforward to show that it descends to the pseudodual of the PCM. (Higher order poles in $\omega$ were considered in \cite{benini2020homotopical}, but not the explicit example appearing here. Since this work first appeared the pseudodual of the PCM was obtained from \CSt{4} using a slightly different, though related, construction in \cite{lacroix2021integrable}.)


\subsection{Trigonometric action} \label{subsec:Trig}

We now derive an alternative action for the ASDYM equations. We will refer to this as the trigonometric action, as the boundary conditions we impose are analogous to those which describe quantum integrable spin chains with trigonometric $R$-matrices \cite{costello2017gauge}. It is closely related to an action of Mason and Sparling \cite{mason1996integrability}.\\

We again consider HCS on $\PT$, but now take
\[ \Omega = \frac{\Diff^3Z}{(Z\cdot A_+)(Z\cdot A_-)(Z\cdot B_+)(Z\cdot B_-)} = \frac{\la\diff\pi\,\pi\ra\wedge\diff^2x^{A'B'}\pi_{A'}\pi_{B'}}{2\la\pi\,\alpha_+\ra\la\pi\,\alpha_-\ra\la\pi\,\beta_+\ra\la\pi\,\beta_-\ra}\,. \]
Note that we are assuming the loci $Z\cdot A_\pm,Z\cdot B_\pm=0$ all intersect along the $\CP^1$ at infinity. We will assume without loss of generality that $\la\alpha_+\,\beta_+\ra = \la\alpha_-\,\beta_-\ra = 1$. Once again, varying the action on the support of the bulk equations of motion generates boundary terms at the locations of the poles.  For example, the unwanted boundary term at $\pi\sim\alpha_+$ is
\be
{\delta S}_\Omega|_{\pi=\alpha_+} = \frac{1}{2\la\alpha_+\,\alpha_-\ra\la\alpha_+\,\beta_-\ra}\int_\E\diff^2x^{A'B'}{\alpha_+}_{A'}{\alpha_+}_{B'}\wedge\tr(\delta{\bar\cA}\wedge{\bar\cA})|_{\pi=\alpha_+}\,. \label{eq:SimPol}
\ee
Requiring that ${\bar\cA}$ be divisible by $\la\pi\,\alpha_+\ra$ would introduce a quadratic zero in $\tr(\delta {\bar\cA}\wedge{\bar\cA})$ at $\pi\sim\alpha_+$. This was needed in section~\ref{sec:sec2} where $\Omega$ had double poles, but here it is unnecessarily strong. In this section we will consider less stringent  boundary conditions, known as trigonometric boundary conditions, that were originally proposed in \cite{costello2017gauge} and first applied in the context of 2d classical integrable field theory in \cite{costello2019gauge}. We briefly review these boundary conditions now.\\

We assume that our Lie algebra $\fg$ is a Manin triple. We will be particularly interested in a class of non-simple Manin triples, and so we relax our assumptions on $\fg$ by allowing it to be reductive. We still assume the existence of an invariant bilinear on $\fg$ which we continue to denote by $\tr$. A Manin triple admits the following decomposition
\[ \fg \cong \fl_-\dotplus\fl_+ \]
for $\fl_\pm$ disjoint Lagrangian subalgebras of $\fg$. Here $\dotplus$ indicates the direct sum as vector spaces.\\

To construct a Manin triple we start by choosing a complex simple Lie algebra $\fg_0$. This comes with an essentially unique $\fg_0$-invariant symmetric bilinear, the Killing form $\tr_0$. Fixing a choice of Cartan subalgebra and base we have the standard decomposition $\fg_0 = \fn_-\dotplus\fh\dotplus\fn_+$. We then define
\[ \fg = \fg_0\oplus\widetilde\fh\,, \]
where $\widetilde \fh$ is a second copy of the Cartan. Note that by $\oplus$ we mean the direct sum as Lie algebras, so $\fg$ is a central extension of $\fg_0$. We define our $\fg$-invariant bilinear by
\[ \tr\big((x+{\widetilde h}_1)(y+{\widetilde h}_2)\big) = \tr_0(xy) - \tr_0({\widetilde h}_1{\widetilde h}_2)\,, \]
and identify
\[ \fl_- = \fn_-\oplus\fh_-\,,\quad\fl_+ = \fn_+\oplus\fh_+\,, \]
where ($\fh_-$) $\fh_+$ is the (anti-)diagonal subgroup of $\fh\oplus{\widetilde \fh}\subset\fg$. It should be clear that these are disjoint Lagrangian subalgebras.\footnote{This is not the most general choice of $\fl_\pm$ in $\fg_0\oplus{\widetilde\fh}$, see \cite{costello2018gauge} for a more generic choice.}\\

We can also perform this decomposition at the level of the Lie group. Let $G_0$ be the complex simple Lie group with Lie algebra $\fg_0$. We write $H$ and $N_\pm$ for its subgroups with Lie algebras $\fh$ and $\fn_\pm$ respectively. We may then identify the dense open set $U = N_-HN_+\subset G_0$ of elements $g_0\in G_0$ which can be expressed in the form $g_0 = n_-hn_+$ for $h\in H$, $n_\pm\in N_\pm$. This coincides with the big Bruhat cell, and within it this decomposition is unique. Define the complex reductive Lie group $G = G_0\times{\widetilde H}$ which has Lie algebra $\fg$. It has subgroups
\[ L_- = N_- H_-\,,\qquad L_+ = H_+ N_+ \]
with Lie algebras $\fl_\pm$ respectively. Here ($H_-$) $H_+$ is the (anti-)diagonal subgroup of $H\times\widetilde H$. The natural map
\[ L_-\times L_+ \to U\times\widetilde H\subset G \]
is a $2^{\mathrm{rank}\,\fg_0}$-fold cover, since inverting it requires taking the square root of an element of $\fh$.\\

Taking $G$ to be the gauge group of HCS, we enforce the following boundary conditions on $\bar\cA$ at $\pi\sim\alpha_\pm,\beta_\pm$ 
\[ \bar\cA|_{\pi\sim\alpha_\pm},\bar\cA|_{\pi\sim\beta_\pm}\in\fl_\mp\,, \]
where the strange choice of signs is for later convenience. These eliminate the boundary terms by virtue of the fact that $\fl_\pm$ are isotropic. We must also have
\[ \varepsilon|_{\pi\sim\alpha_\pm},\varepsilon|_{\pi\sim\beta_\pm}\in\fl_\mp \]
if infinitesimal gauge transformations parametrised by $\varepsilon$ are to preserve these boundary conditions.\footnote{A minor technicality: we should really demand that the projection of $\bar\cA$, and of an infinitesimal gauge transformation $\varepsilon$, onto the subalgebra $\fl_\pm$ is divisible by $\la\pi\,\alpha_\pm\ra\la\pi\,\beta_\pm\ra$.}\\

We now wish to determine the effective space-time theory. As usual, we begin by pulling back the gauge field to the twistor fibres and expressing it in terms of $\sigha$ as
\[ {\bar\cA}_x = \iota^*_x{\bar\cA} = \sigha^{-1}{\bar\partial}_{\CP^1_x}\sigha\,. \]
Gauge transformations $g:\PT\to G$, which act on $\sigha$ by $\sigha\mapsto\sigha g^{-1}$, are arbitrary away from $\pi\sim\alpha_\pm,\beta_\pm$. $\sigha$ is determined up to gauge by its values at these points. Introducing $\sigma_{\alpha_\pm} = \sigha|_{\pi\sim\alpha_\pm}$ and $\sigma_{\beta_\pm} = \sigha|_{\pi\sim\beta_\pm}$, under a gauge transformation $\sigma_{\alpha_\pm}\mapsto\sigma_{\alpha_\pm}g^{-1}|_{\pi\sim\alpha_\pm}$ and $\sigma_{\beta_\pm}\mapsto\sigma_{\beta_\pm}g^{-1}|_{\pi\sim\beta_\pm}$ for $g|_{\pi\sim\alpha_\pm},g|_{\pi\sim\beta_\pm}:\E^4\to L_\mp$. We also have the redundancy $\sigha\mapsto h\sigha$ for $h:\E^4\to G$. Therefore up to gauge $\sigha$ is determined by a map
\[ \sigma = (\sigma_{\alpha_+},\sigma_{\alpha_-},\sigma_{\beta_+},\sigma_{\beta_-}):\E^4\to G\backslash(G/L_-\times G/L_+\times G/L_-\times G/L_+)\,. \]
We can understand the target space better by fixing the redundancies. First use the action on the left by $G$ to fix $\sigma_{\beta_+}=\id$. This leaves the freedom to act on the left with elements of $L_-$, since these can be compensated by the action on the right of $\sigma_{\beta_+}$. Assuming that $\sigma_{\beta_-}$ takes values in $U$, we can write it as a product $\sigma_{\beta_-} = \ell_-^{-1}\ell_+$ for $\ell_\pm\in L_\pm$. Then acting on the left with $\ell_-$ and on the right with $\ell_+^{-1}$ we can also set $\sigma_{\beta_-}=\id$. This exhausts the freedom to act on the left. Assuming that $\sigma_\pm = \sigma_{\alpha_\pm}$ take values in $U$, we can fix the remaining right actions by translating them into $L_\pm$ respectively. Therefore $\sigha$ is determined by a pair of maps $\sigma_\pm:\E^4\to L_\pm$.\\

Next, we turn our attention to the remaining components of $\bar\cA$. Since $\Omega$ is nowhere vanishing we have
\[ \bar\cA = \sigha^{-1}{\bar\partial}\sigha + \sigha^{-1}\bar\cA'\sigha\,, \]
for $\bar\cA' = {\hat e}^A\pi^{A'}A_{AA'}$ with $A$ independent of $\pi$. Our boundary conditions then fix
\[ A_{AA'} =  -{\beta_+}_{A'}{\alpha_+}^{B'}\p_{AB'}\sigma_+\sigma_+^{-1} - {\beta_-}_{A'}{\alpha_-}^{B'}\p_{AB'}\sigma_-\sigma_-^{-1}\,. \]
Note that a general ASD connection can always be written in this form. To achieve this one first projects onto $\fl_\pm$, and then introduces a Yang matrix $\sigma_\pm$ with values in $L_{\pm}$ for each projection.\\

Substituting this expression for $\bar\cA$ back into the action we obtain an effective action in terms of $\sigma_\pm$. This computation is essentially identical to the one performed when deriving the LMP action, so we omit the details here. It gives
\be \label{eq:STrig}
S_\mathrm{Trig}[\sigma_\pm] = \frac{1}{\la\alpha_+\,\alpha_-\ra}\int_{\E^4}\mathrm{vol}_\delta\,\varepsilon^{AB}\alpha_-^{A'}\alpha_+^{B'}\tr({J_-}_{AA'}{J_+}_{BB'})\,,
\ee
where $J_- = -\diff\sigma_-\sigma_-^{-1}$ and  $J_+ = -\diff\sigma_+\sigma_+^{-1}$. Introducing
\be \p = \frac{\diff x^{AA'}\alpha_{+A'}\alpha_-^{B'}\p_{AB'}}{\la\alpha_-\,\alpha_+\ra}\,,\qquad \tilde\p = \frac{\diff x^{AA'}\alpha_{-A'}\alpha_+^{B'}\p_{AB'}}{\la\alpha_+\,\alpha_-\ra}\,. \ee
we can rewrite the action
\[
S_\mathrm{Trig}[\sigma_\pm] = \frac{1}{\la\alpha_-\,\alpha_+\ra}\int_{\E^4}\mu_{\alpha_-,\alpha_+}\wedge\tr\big(\p\sigma_-\sigma_-^{-1}\,\tilde\p\sigma_+\sigma_+^{-1}\big)\,.
\]
If $\alpha_+\sim\hat\alpha_-$ then $\p,\tilde\p$ coincide with the Dolbeault operators, and $\mu_{\alpha_-,\hat\alpha_-}$ with the K\"{a}hler form (up to an overall factor), in the complex structure on space-time determined by $\alpha_-$.\\

The corresponding classical equations of motion are
\[ \big[\mu_{\alpha_-,\alpha_+}\wedge\partial(\sigma_-^{-1}{\tilde\p}\sigma_+\sigma_+^{-1}\sigma_-)\big]_{\fl_+} = 0\,,\quad \big[\mu_{\alpha_-,\alpha_+}\wedge{\tilde\p}(\sigma_+^{-1}{\partial}\sigma_-\sigma_-^{-1}\sigma_+)\big]_{\fl_-} = 0\,.
\]
Here the square brackets indicate that we are projecting onto the relevant subalgebra. These equations are in fact equivalent to the ASDYM equations for a $\fg$-valued gauge field. A proof of this is included in appendix \ref{app:trigASDYM}.\\

We can also understand these equations of motion in terms of fields taking values in the simple Lie group $G_0$. To do so we write
\[ \sigma_-^{-1} = (\ell h_-,h_-^{-1})\,,\quad \sigma_+ = (h_+u,h_+)\,, \]
where $h_\pm$ take values in $H$, and $u$, $\ell$ take values in $N_\pm$ respectively. Projecting Yang's equation onto ${\widetilde \fh}$, which is abelian, we find that
\[ \mu_{\alpha_-,\alpha_+}\wedge\p({\tilde\p}(h_-^{-1}h_+)h_+^{-1}h_-) = \mu_{\alpha_-,\alpha_+}\wedge\partial({\tilde\p}h_+h_+^{-1} - {\tilde\p}h_-h_-^{-1}) = 0\,. \]
Under the assumption that $h_\pm\to{\rm id.}$ as $\|x\|\to\infty$ in $\E^4$, the solution to this equation is $h_+=h_-=h$. We are left with Yang's equation for $\ell h^2 u$. \\

Imposing $h_\pm=h$ at the level of the action we obtain
\be \label{eq:SMasSpa} S_\mathrm{MS}[L,U] = \frac{1}{\la\alpha_-\,\alpha_+\ra}\int_{\E^4}\mu_{\alpha_-,\alpha_+}\wedge\tr_0\bigg(\p LL^{-1}\wedge{\tilde\p} U U^{-1} - \frac{1}{2}\p LL^{-1}\wedge{\tilde\p}LL^{-1}\bigg) \ee
where $u=U\in N_+$ and $L = h^{-2}\ell^{-1}\in B_-=N_-H$. A derivation is included in appendix \ref{app:trigASDYM}. This action for the ASDYM equations was originally proposed by Mason and Sparling in \cite{mason1996integrability}.\\

Unfortunately the actions \eqref{eq:STrig} and \eqref{eq:SMasSpa} cannot be used to obtain any new 2d integrable field theories via symmetry reduction. This is because they agree with the action of \WZWt{4} with gauge group $G$ and $G_0$ respectively. This can be seen by substituting $\sigma = L^{-1}U$ and $\sigma = \sigma_-^{-1}\sigma_+$ into equation \ref{eq:WZW4} and applying the Polyakov-Wiegmann identity. Curiously this agreement is not manifest on twistor space.\\

As such performing a symmetry reduction by a 2d group of translations on $\E^4$ reproduces the results of section \ref{sec:sec2}, and we will obtain the PCM. These actions will, however, be trigonometric in character. Indeed quotienting \eqref{eq:STrig} by the 2d group of translations generated by $X = \hat\kappa^A\hat\gamma^{A'}\p_{AA'}$, $\bar X = \kappa^A\gamma^{A'}\p_{AA'}$ we obtain
\be \label{eq:SPCMTrig} S_\mathrm{PCMTrig}[\sigma_\pm] \propto \int_\Sigma\bigg(\frac{\la\gamma\,\alpha_+\ra}{\la\gamma\,\alpha_-\ra}\tr(J^{1,0}_+\wedge J^{0,1}_-) - \frac{\la\hat\gamma\,\alpha_+\ra}{\la\hat\gamma\,\alpha_-\ra}\tr(J^{1,0}_-\wedge J^{0,1}_+)\bigg) \ee
for $\sigma_\pm:\Sigma\to L_\pm$ and $J_\pm = -\diff_\Sigma\sigma_\pm\sigma_\pm^{-1}$. This is indeed equivalent to the PCM derived in subsection \ref{subsec:SymRedE}, as can be seen by making the substitution $\sigma = \sigma_-^{-1}\sigma_+$ in equation \eqref{eq:PCM} and applying the Polyakov-Wiegmann identity. Alternatively we could have performed this reduction directly on twistor space. Doing so we'd obtain \CSt{4} on $\cV= \Sigma\times\CP^1$ with
\[ \omega = \frac{\la\pi\,\gamma\ra\la\pi\,\hat\gamma\ra\la\diff\pi\,\pi\ra}{\la\pi\,\alpha_+\ra\la\pi\,\alpha_-\ra\la\pi\,\beta_+\ra\la\pi\,\beta_-\ra}\,. \]
The boundary conditions are $A'|_{\pi\sim\alpha_\pm},A'|_{\pi\sim\beta_\pm}\in\fl_\mp$, and we allow a simple pole in $A'_w, A'_{\bar w}$ at $\pi=\gamma,\hat\gamma$ respectively. In \cite{costello2019gauge} it was demonstrated that \CSt{4} for this choice of $\omega$ was equivalent to \eqref{eq:SPCMTrig}. (The condition that $\gamma$ and $\hat\gamma$ be antipodal is an artefact of working in Euclidean signature. See section \ref{sec:ReSt} for how to bypass this constraint.) \\

In \cite{delduc2020unifying} generalisations of the trigonometric boundary conditions appearing in \cite{costello2019gauge} were studied in the context of \CSt{4}. We anticipate that introducing analogues of these boundary conditions on twistor space at simple poles in $\Omega$ will allow one to obtain 4d analogues of the Yang-Baxter $\sigma$-model and $\lambda$-deformed PCM. (Investigations of this nature have since appeared in \cite{chen2021deformed}.)


\subsection{4d integrable coupled $\sigma$-models} \label{subsec:CoSi}

So far we have only considered $\Omega$ which are nowhere vanishing. We now briefly consider what happens if we relax this constraint. The effective space-time actions we obtain do not have classical equations of motion equivalent to the ASDYM equations. They do, however, admit a 4d Lax connection, and we will later see their symmetry reductions describe known 2d integrable theories. Furthermore, it is clear from the twistor description that the equations of motion for the models we obtain are related to holomorphic bundles over twistor space, albeit with the caveat that the partial connection may admit poles.\\

We restrict ourselves to the following, fairly general, choice of $\Omega$
\[ \Omega = \frac{\Diff^3Z\prod_{j=1}^n(Z\cdot A_j)(Z\cdot B_j)}{\prod_{i=1}^{n+2}(Z\cdot C_i)^2} = \frac{\la\diff\pi\,\pi\ra\diff^2x^{A'B'}\pi_{A'}\pi_{B'}\prod_{j=1}^n\la\pi\,\alpha_j\ra\la\pi\,\beta_j\ra}{2\prod_{i=1}^{n+2}\la\pi\,\gamma_i\ra^2}\,. \]
For the second equality to hold the loci $Z\cdot A_j,Z\cdot B_j,Z\cdot C_i=0$ must all intersect along the $\CP^1$ at infinity. We introduce the standard boundary conditions for $\bar\cA$ at the double poles of $\Omega$, {\it i.e.} that $\bar\cA$ is divisible by $\prod_{i=1}^{n+2}\la\pi\,\gamma_i\ra$. At zeros of $\Omega$ we can tolerate simple poles in components of $\hat\cA_A$ without generating boundary terms when varying the action. More precisely, introducing the right-handed dyad $[\mu\,\nu]=1$, we permit simple poles in $\mu^A{\hat\cA}_A$ and $\nu^A{\hat\cA}_A$ at $\pi\sim\alpha_j$ and $\pi\sim\beta_j$ respectively for all $j=1,\dots,n$. Note that the combination $\Omega\wedge\tr(\delta{\bar\cA}\wedge\bar\cA)$, which appears when varying the action on the support of the equations of motion, is free from poles.\\

The pullback of ${\bar\cA}$ to the twistor fibres, ${\bar\cA}_x = \iota^*_x{\bar\cA}$, has no poles. We introduce $\sigha:\PT\to G$ in the usual way
\[ {\bar\cA}_x = \sigha^{-1}{\bar\partial}_x\sigha\,. \]
The gauge invariant data that can be extracted from ${\bar\cA}_x$ is furnished by the holomorphic Wilson lines between the $\gamma_i$, and can be completely characterised by the map
\[ \sigma:\E^4\to G\backslash G^{n+2}\,,\quad x\mapsto [(\sigma_1,\dots,\sigma_{n+2})]= [(h\sigma_1,\dots,h\sigma_{n+2})] \]
with $\sigma_i = \hat\sigma|_{\gamma_i}$. We could identify $G\backslash G^{n+2}$ with $G^{n+1}$ by fixing $\sigma_{n+2}={\rm id.}$, but we will find it convenient not to do so. We assume $\sigha$ to be take the value $\sigma_i$ in a neighbourhood of $\gamma_i$ for all $i=1,\dots,n+2$. Then the gauge invariant holomorphic Wilson lines are
\[ \cW_{\gamma_i\to\gamma_j} = \sigha^{-1}|_{\gamma_j}\sigha|_{\gamma_i} = \sigma_j^{-1}\sigma_i\,. \]
The redundancy $(\sigma_1,\dots,\sigma_{n+2})\mapsto(h\sigma_1,\dots,h\sigma_{n+2})$ will be gauge symmetry of the resulting 4d theory.\\

The next step is to solve the classical equations of motion in the directions of the fibres. We have
\[ {\bar\cA} = \sigha^{-1}{\bar\partial}\sigha + \sigha^{-1}{\bar\cA}'\sigha \]
where ${\hat\cA}_A'$ is meromorphic in $\pi$. Simple poles are permitted at the $\alpha_j$ in $\mu^A{\hat\cA_A}$ and at the $\beta_j$ in $\nu^A{\hat\cA_A}$. Our boundary conditions imply $\bar\cA|_{\pi\sim\gamma_i}=0$ for $i=1\dots n+2$, hence
\[  \hat\cA'_A|_{\pi\sim\gamma_i} = - \frac{\la\pi\,{\hat\gamma}_i\ra\gamma^{A'}_i\partial_{AA'}\sigma_i\sigma_i^{-1}}{\|\gamma_i\|^2}\bigg\rvert_{\pi\sim\gamma_i}\,. \]
The unique choice for $\bar\cA'$ obeying these constraints is
\be {\hat\cA}_A' = \sum_{i=1}^{n+2}\prod_{j=1,\,j\neq i}^{n+2}\frac{\la\pi\,\gamma_j\ra}{\la\gamma_i\,\gamma_j\ra}\bigg(\nu_A\mu^B\prod_{k=1}^n\frac{\la\gamma_i\,\alpha_k\ra}{\la\pi\,\alpha_k\ra} - \mu_A\nu^{B}\prod_{k=1}^n\frac{\la\gamma_i\,\beta_k\ra}{\la\pi\,\beta_k\ra}\bigg) \gamma^{B'}_iJ_{iBB'}\,, \label{eq:Laxmxdef} \ee
where we've defined $J_i = - \diff\sigma_i\sigma_i^{-1}$. Note that it is essential that we allow simple poles in $\bar\cA$ in order to satisfy the boundary conditions. We also emphasise that $\bar\cA'$ is not linear in $\pi_{A'}$ for $n>0$, and so cannot be straightforwardly related to a space-time gauge field. We can, however, still interpret ${\hat\cA}_A' = {\hat\scrL}_A$ as a 4d Lax connection.\\

Under the redundancy $(\sigma_1,\dots,\sigma_{n+2})\mapsto(h\sigma_1,\dots,h\sigma_{n+2})$ we have
\[ {\hat\scrL}_A\mapsto h{\hat\scrL}_A h^{-1} - \sum_{i=1}^{n+2}\prod_{j=1,\,j\neq i}^{n+2}\frac{\la\pi\,\gamma_j\ra}{\la\gamma_i\,\gamma_j\ra}\bigg(\nu_A\mu^B\prod_{k=1}^n\frac{\la\gamma_i\,\alpha_k\ra}{\la\pi\,\alpha_k\ra} - \mu_A\nu^{B}\prod_{k=1}^n\frac{\la\gamma_i\,\beta_k\ra}{\la\pi\,\beta_k\ra}\bigg) \gamma^{B'}_i\partial_{BB'}hh^{-1}\,. \]
To simplify this we use
\[ \sum_{i=1}^{n+2}\prod_{j=1,\,j\neq i}^{n+2}\frac{\la\pi\,\gamma_j\ra}{\la\gamma_i\,\gamma_j\ra}\prod_{k=1}^n\frac{\la\gamma_i\,\alpha_k\ra}{\la\pi\,\alpha_k\ra}\gamma_i^{A'} = \pi^{A'}\,, \]
which can be verified by evaluating both sides at $\pi\sim\alpha_i$, and computing the residues at $\pi\sim\alpha_k$. The same identity holds if we replace $\alpha_j$ by $\beta_j$. We therefore deduce that
\[ {\hat\scrL}_A \mapsto h{\hat\scrL}_Ah^{-1} - \pi^{A'}\partial_{AA'}hh^{-1} \]
as expected.\\

Following by now fairly standard methods we can determine the effective space-time action. It is given by
\[ \begin{aligned}
&\sum_{i=1}^{n+2}a_ib_i\Bigg(\frac{1}{2}\int_{\E^4}\tr(J_i\wedge\ast_4J_i) - \sum_{k=1}^n\frac{\la\alpha_k\,\beta_k\ra}{\la\alpha_k\,\gamma_i\ra\la\gamma_i\,\beta_k\ra}\int_{\E^4}\mathrm{vol}_\delta\,\mu^{(A}\nu^{B)}\gamma_i^{A'}\gamma_i^{B'}\tr({J_i}_{AA'}{J_i}_{BB'}) \\
&\quad+ \frac{1}{3}\int_{\E^4\times[0,1]}\diff^2x^{A'B'}{\gamma_i}_{A'}\Bigg(\sum_{j=1\,j\neq i}^{n+2}\frac{{\gamma_j}_{B'}}{\la\gamma_i\,\gamma_j\ra} - \frac{1}{2}\sum_{k=1}^n\bigg(\frac{{\alpha_k}_{B'}}{\la\gamma_i\,\alpha_k\ra} + \frac{{\beta_k}_{B'}}{\la\gamma_i\,\beta_k\ra}\bigg)\Bigg)\wedge\tr({\tilde J}_i^3)\Bigg) \\
&\quad\quad+ \sum_{i,j=1\,i\neq j}^{n+2}\frac{1}{\la\gamma_i\,\gamma_j\ra}\int_{\E^4}\mathrm{vol}_\delta\,\bigg(a_ib_j\mu^A\nu^B - a_jb_i\mu^B\nu^A\bigg)\gamma_i^{A'}\gamma_j^{B'}\tr({J_i}_{AA'}{J_j}_{BB'})\,.
\end{aligned} \]
Here we have introduced
\[ a_i = \frac{\prod_{k=1}^n\la\gamma_i\,\alpha_k\ra}{\prod_{j=1,\,j\neq i}^{n+2}\la\gamma_i\,\gamma_j\ra}\,,\quad b_i = \frac{\prod_{k=1}^n\la\gamma_i\,\beta_k\ra}{\prod_{j=1,\,j\neq i}^{n+2}\la\gamma_i\,\gamma_j\ra}\,.
\]
$\tilde\sigma$ is a smooth homotopy from $\sigma$ to the diagonal in $G\backslash G^{n+2}$, and as usual ${\tilde J}_i = -{\tilde\diff}{\tilde\sigma}_i{\tilde\sigma}_i^{-1}$. There are a couple of sanity checks we can perform to confirm to test whether this is the correct action.
\begin{itemize}
	\item The first is that it's invariant under the gauge transformation $(\sigma_1,\dots,\sigma_{n+2})\mapsto (h\sigma_1,\dots,h\sigma_{n+2})$. Infinitesimally we have
	\[ \delta\sigma_i = \varepsilon\sigma_i \]
	and so
	\[ \delta J_i = -\delta(\diff\sigma_i\sigma_i^{-1}) = -\diff\varepsilon - [\varepsilon,\diff\sigma_i\sigma_i^{-1}] = -\diff\varepsilon - [\varepsilon,J_i] \,. \]
	It is clear that the action is invariant under the simultaneous adjoint action on the currents $J_i$, and so it is enough to establish invariance under
	\[ \delta J_i = -\diff\epsilon\,. \]
	We find that\footnote{In deriving this we have made use of the spinor identities
		\[ \frac{{\hat\gamma_i}^{A'}}{\|\gamma_i\|^2} - \bigg(\sum_{j=1,\,j\neq i}^{n+2}\frac{\la\gamma_j\,{\hat\gamma}_i\ra}{\la\gamma_j\,\gamma_i\ra} - \sum_{k=1}^n\frac{\la\alpha_k\,{\hat\gamma}_i\ra}{\la\alpha_k\,\gamma_i\ra}\bigg)\frac{\gamma_i^{A'}}{\|\gamma_i\|^2} = \sum_{j=1,\,j\neq i}^{n+2}\frac{\gamma_j^{A'}}{\la\gamma_i\,\gamma_j\ra} - \sum_{k=1}^n\frac{\alpha_k^{A'}}{\la\gamma_i\,\alpha_k\ra}\,, \]
		the same identity with all $\alpha_j$ are replaced with $\beta_j$, and
		\[ \sum_{k=1}^n\frac{\la\alpha_k\,\beta_k\ra}{\la\alpha_k\,\gamma_i\ra\la\gamma_i\,\beta_k\ra} =  \frac{1}{\|\gamma_i\|^2}\bigg(\sum_{k=1}^n\frac{\la\alpha_k\,{\hat\gamma}_i\ra}{\la\alpha_k\,\gamma_i\ra} - \sum_{k=1}^n\frac{\la\beta_k\,{\hat\gamma}_i\ra}{\la\beta_k\,\gamma_i\ra}\bigg)\,. \]}
	\[ \begin{aligned} \delta S &= 2\sum_{i=1}^na_ib_i\int_{\E^4}\mathrm{vol}_\delta\,\,\tr(\partial_{AA'}\varepsilon {J_i}_{BB'})\gamma_i^{B'}\Bigg(\mu^A\nu^B\Bigg(\sum_{j=1,\,j\neq i}^{n+2}\frac{\gamma_j^{A'}}{\la\gamma_i\,\gamma_j\ra}\bigg(1+\frac{a_j}{a_i}\bigg) \\ &\phantom{{}={}} \negmedspace- \sum_{k=1}^n\frac{\alpha_k^{A'}}{\la\gamma_i\,\alpha_k\ra}\Bigg) - \mu^B\nu^A\Bigg(\sum_{j=1,\,j\neq i}^{n+2}\frac{\gamma_j^{A'}}{\la\gamma_i\,\gamma_j\ra}\bigg(1+\frac{b_j}{b_i}\bigg) - \sum_{k=1}^n\frac{\beta_k^{A'}}{\la\gamma_i\,\beta_k\ra}\Bigg)\Bigg)\,.
	\end{aligned} \]
	This vanishes since
	\[ \sum_{j=1,\,j\neq i}^{n+2}\frac{\gamma_j^{A'}}{\la\gamma_i\,\gamma_j\ra}\bigg(1+\frac{a_j}{a_i}\bigg) = \sum_{k=1}^n\frac{\alpha_k^{A'}}{\la\gamma_i\,\alpha_k\ra}\,, \]
	and similarly if we replace $\alpha_j$ with $\beta_j$.
	\item The second is that taking the limit $\alpha_n,\beta_n\to\gamma_{n+2}$ effectively reduces $n$ to $n-1$. This property is manifest for the twistor action. On space-time it follows from the fact that in this limit $a_{n+2} = b_{n+2} = 0$, and
	\[ \frac{\gamma_{n+2}^{A'}}{\la\gamma_i\,\gamma_{n+2}\ra} = \frac{\alpha_n^{A'}}{\la\gamma_i\,\alpha_n\ra} = \frac{\beta_n^{A'}}{\la\gamma_i\,\beta_n\ra}\,.\]
\end{itemize}
The classical equations of motion are
\be \sum_{j=1,\,j\neq i}^{n+2}\frac{\gamma_i^{A'}\gamma_j^{B'}}{\la\gamma_i\,\gamma_j\ra}(a_ib_j\mu^A\nu^B - a_jb_i\mu^B\nu^A)\big(\partial_{AA'}{J_j}_{BB'} - \partial_{BB'}{J_i}_{AA'} + [{J_i}_{AA'},{J_j}_{BB'}]\big) = 0 \label{eq:coupledeom}\ee
for $i=1,\dots,n+2$. Note that the gauge symmetry is manifest.\\

We can understand these equations of motion by working directly with the Lax equation
\be {\hat\partial}_A{\hat\scrL}^A + \frac{1}{2}[{\hat\scrL}_A,{\hat\scrL}^A] = 0\,.
\label{eq:lax}
\ee
We will assume that $\scrL$ is meromorphic in $\pi$, and that $\mu^A{\hat\scrL}_A$ and $\nu^A{\hat\scrL}_B$ have simple poles at $\pi = \alpha_j$ and $\pi = \beta_j$ respectively for $j=1,\dots,n$. We can solve this equation at any $\pi$ away from the simple poles by expressing ${\hat\scrL}_A$ in pure gauge. Doing so at $\gamma_i$ for $i=1,\dots,n+2$ we have
\be \label{eq:laxgammai} \hat\scrL_A|_{\pi\sim\gamma_i} = -\hat\p_A\sigma_i\sigma_i^{-1}|_{\pi\sim\gamma_i} \ee
where $\sigma:\E\to G^n$. This is enough to completely determine $\hat\scrL_A$ in terms of $\sigma$. Indeed this reproduces equation \eqref{eq:Laxmxdef}.\\

The Lax equation is invariant under the standard gauge symmetry, ${\hat\scrL}_A \mapsto h{\hat\scrL}_Ah^{-1} - {\hat\partial}_Ahh^{-1}$, for $h$ independent of $\pi$, which we have seen acts on $\sigma$ by mapping $\sigma_i\mapsto h\sigma_i$. Hence modulo this gauge symmetry $\sigma$ takes values in $G\backslash G^{n+2}$.\\

Requiring that the lax equation holds to second order at the $\gamma_i$, which is achieved by taking the Lie derivative of \eqref{eq:lax} along
\[ \xi_i=-\frac{\la\pi\,{\hat\gamma}_i\ra{\hat\gamma}_i^{A'}}{\|\gamma_i\|^2}\frac{\p}{\p\pi^{A'}}
\]
and evaluating at $\pi\sim\gamma_i$, gives the equations of motion \eqref{eq:coupledeom}.\\

Conversely the conditions \eqref{eq:laxgammai} and equations of motion \eqref{eq:coupledeom} are sufficient to ensure that the Lax equation holds for all $\pi$. To see why, note that the left hand side of \eqref{eq:lax} is a meromorphic section of $\cO(2)$ with simple poles at $\pi\sim\alpha_i,\beta_i$. This is a $2n+3$ dimensional space. The conditions \eqref{eq:laxgammai} ensure that it vanishes to first order at the $\gamma_i$, and the conditions \eqref{eq:coupledeom} extend this to second order. This gives $2n+4$ constraints in total, with one lost to the gauge symmetry. The Lax equation follows.\\

Unfortunately we have been unable to give a straightforward interpretation to the Lax equation on space-time. It is our expectation that this deformed theory, and similar deformations will be in appropriate sense integrable classical field theories. We will leave a more thorough analysis for future work.\\

Finally we consider the result of performing a symmetry reduction by a 2d group of translations to this theory. It is easiest to understand this reduction directly on $\PT$. Taking $\nu = \hat\mu$, and $\beta_j = \hat\alpha_j$, quotienting by the 2d group of translations generated by $\cX = \hat\mu^A\hat\alpha_{n+1}^{A'}\p_{AA'}$, $\bar\cX = \mu^A\alpha_{n+1}^{A'}\p_{AA'}$ gives \CSt{4} on $\cV = \E^2\times\CP^1$ with
\[ \omega = \frac{\prod_{j=1}^{n+1}\la\pi\,\alpha_j\ra\la\pi\,\hat\alpha_j\ra\la\diff\pi\,\pi\ra}{\prod_{i=1}^{n+2}\la\pi\,\gamma_i\ra^2}\,. \]
$A'$ has the standard boundary conditions at the quadratic poles in $\omega$, and we tolerate simple poles in $A_w'$ and $A_{\bar w}'$ at $\pi\sim\alpha_j$ and $\pi\sim\hat\alpha_j$ respectively for $j=1,\dots,n$. (That the simple zeros in $\omega$ lie at antipodal points is an artefact of working in Euclidean signature. We will see shortly how to bypass this restriction.) We recognise this as the \CSt{4} realisation~\cite{costello2019gauge} of the theory of integrable coupled 2d $\sigma$-models introduced in~\cite{delduc2019integrable}. It is straightforward, if tedious, to verify that performing the reduction on $\E^4$ leads to the same theory.\\

The equations of motion this theory of integrable coupled 2d $\sigma$-models do not appear to arise as a symmetry reduction of the ASDYM equations. We have seen here that they do, however, arise as a symmetry reduction of HCS on twistor space. It would be interesting to explore which other integrable systems arise as symmetry reductions of HCS but not as reductions of the ASDYM equations.


\section{Reality conditions} \label{sec:ReSt}

In this section we generalise the results of the preceding chapters to Lorentzian and ultrahyperbolic signatures. This is important because many lower dimensional integrable systems are known to arise from reductions of the ASDYM equations only in ultrahyperbolic signature. We also discuss the related issue of how to restrict the gauge group to a real form.


\subsection{Lorentzian and ultrahyperbolic signatures}

So far we have studied HCS on the twistor space of Euclidean space-time. At first glance the theory does not straightforwardly generalize to Lorentzian and ultrahyperbolic signatures. For example, the twistor space of Minkowksi space has only 5 real dimensions, and does not fibre over space-time. Instead of attempting to define HCS theory on twistor space, it is more fruitful to realise it on the left-handed projective spin bundle. In Euclidean signature this coincides with $\PT$, so this is no different. In other signatures, the projective spin bundle fibres over both space-time $\cM$ and its associated twistor space $\cPT$. In this context, it is often called the correspondence space and we will denote it by $\cF$. As a smooth manifold $\cF\cong\cM\times\CP^1$. A review of the twistor correspondence in arbitrary signature is included in appendix \ref{app:MUtwistors}.\\

Apart from in Euclidean signature, the correspondence space is not naturally a complex manifold. To overcome this, first note that HCS on $\PT$ is equivalent to the following action
\be \frac{1}{2\pi i}\int_\PT\Omega\wedge\mathrm{CS}(\cA) \label{eq:fullHCS}\,, \ee
where $\mathrm{CS}(\cA)$ is the full CS $3$-form constructed using a full connection 1-form $\cA$. Viewing the dynamical field as a full connection means that, in addition to the standard gauge invariance, the action \eqref{eq:fullHCS} has a new, rather trivial, redundancy
\[ \cA \mapsto \cA + \la\diff\pi\,\pi\ra\delta\cA_0 + \diff x^{AA'}\pi_{A'}\delta\cA_A \]
This new gauge freedom can fixed by requiring that $\cA$ is a partial connection, and doing so recovers the standard HCS action. From this perspective it's clear that the theory is only sensitive to the complex structure through the choice of $(3,0)$-form $\Omega$.\\

Fortunately, there is a natural weighted $3$-form on the projective spin bundle over space-time of any signature. Let
\[ \cPT\xleftarrow{P}\cF\xrightarrow{\Pi}\cPT \]
be the twistor correspondence for a real form $\cM$ of complexified space-time $\C\M^4=\C^4$. We can pullback the $(3,0)$-form $\D^3Z$ with holomorphic weight 4 from $\PT$, which we recall is the twistor space of complexified space-time, by the embedding $\iota_\cPT:\cPT\hookrightarrow\PT$. Pulling back again by $P:\cF\to\cPT$ to the correspondence space gives
\[ (\iota_\cPT\circ P)^*\D^3Z = \frac{\la\diff\pi\,\pi\ra\wedge\diff^2x^{A'B'}\pi_{A'}\pi_{B'}}{2}\,. \]
Here we are using coordinates $(x,\pi)$ on $\cF$. We can therefore use the action \eqref{eq:fullHCS} in both Lorentzian and hyperbolic signatures with essentially no modification by treating it as an action on the correspondence space
\be \label{eq:genericCS} \frac{1}{2\pi i}\int_\cF\Omega\wedge\mathrm{CS}(\cA)\,, \ee
with $\Omega = (\iota\circ P)^*(\Phi\D^3Z)$ for $\Phi$ a meromorphic section of $\cO(-4)\to\PT$.\\

It is instructive the compare this situation to that of \CSt{4}. The similarity is uncanny. It too is defined on a space $\cV$ which double fibres over a space-time, $\Sigma$, and a complex manifold, $C$. Furthermore, its Lagrangian is also the wedge product of a top holomorphic form on this complex manifold pulled back to $\cV$ with the CS $3$-form.\\

In sections \ref{sec:sec2} \& \ref{sec:sec3} we demonstrated that in a number of cases that HCS theory on $\PT$ had an effective description on $\E^4$. In fact none of these calculations were sensitive to our choice of real slice $\E^4\subset\C\M^4$. We could equally have started with the action \eqref{eq:genericCS} on $\cF$ over any real form $\cM\subset\C\M^4$. This is reflected in the fact that all of the effective space-time actions we obtain are independent of spinor conjugation in Euclidean signature, $\pi\mapsto\hat\pi$. This may be surprising, since we often made use of spinor conjugation in deriving effective space-time actions, but in fact it was only ever used in fixing the gauge.


\subsection{Real forms of the gauge group} \label{subsec:ReGp}

So far we have understood how to obtain actions for ASDYM equations on space-times of arbitrary signature for a simple, complex gauge group $G$. We now show how to restrict the gauge group to a real form, $G_\R$.\\

Such real forms arise as the fixed point set of an involutive automorphism, $\Theta:G\to G$, which is conjugate-linear on the Lie algebra $\fg$. We write $\theta:\fg\to\fg$ for the induced map on the Lie algebra. The simplest example is $G_\R = \SU_n\subset G = \SL_n(\C)$, for which we may take
\[ \Theta:U\mapsto (U^\dag)^{-1}\,,\qquad\theta:X\mapsto -X^\dag\,. \]
In Lorentzian signature spinor conjugation swaps left- and right-handed spinors, and so swaps the SD and ASD parts of the curvature. As such there are no ASD connections on Minkowski space for real gauge groups. We therefore restrict our attention to Euclidean and ultrahyperbolic signatures.


\subsubsection*{Euclidean signature} \label{subsec:Esig}

In Euclidean signature we can work with a partial connection $\bar\cA\in\Omega^{0,1}(\PT,\fg)$. Writing $C:(x,\pi)\mapsto (x,\hat\pi)$ for spinor conjugation on $\PT$, we impose the reality condition
\be C^*\bar\cA = \theta(\bar\cA)\,. \label{eq:realityconditionAE} \ee
We emphasise that this equation makes sense for a partial connection $\bar\cA$, since both sides of the above equation are $(1,0)$-forms with values in $\fg$. For a gauge transformation with parameter $g$ to preserve this constraint it must obey
\[ C^*g = \Theta(g)\,. \]
Note the curious fact that, since $C$ has no fixed points, at no point in $\PT$ are any components of $\bar\cA$ required to take values in $\fg_\R$, or are gauge transformations required to take values in $G_\R$.

Using the identity
\[ \overline{\HCS(\bar\cA)} = \overline{\HCS}(\theta(\bar\cA))\,, \]
where $\overline\HCS$ is defined using $\p$ instead of $\bar\p$, and taking care to note that $C$ is orientation reversing, we have
\[ \begin{aligned}
S &= \frac{1}{2\pi i}\int\Omega\wedge\HCS(\bar\cA) = -\frac{1}{2\pi i}\int C^*\big(\Omega\wedge\HCS(\bar\cA)\big) = -\frac{1}{2\pi i}\int (C^*\Omega)\wedge\overline\HCS(C^*\bar\cA) \\
&= -\frac{1}{2\pi i}\int(C^*\Omega)\wedge\overline\HCS(\theta({\bar\cA})) = \overline{\frac{1}{2\pi i}\int\overline{(C^*\Omega)}\HCS(\bar\cA)}\,.
\end{aligned} \]
Hence the action will be real if
\[ \overline{C^*\Omega} = \Omega\,. \]
Of course, our boundary conditions on $\bar\cA$ must also be consistent with the reality conditions \eqref{eq:realityconditionAE}.\\

The simplest choice of $\Omega$ compatible with this constraint is a specialisation of the one proposed by Costello and discussed in section \ref{subsec:HCStoWZW4}
\[ \Omega = \frac{\la\diff\pi\,\pi\ra\wedge\diff^2x^{A'B'}\pi_{A'}\pi_{B'}}{2\la\pi\,\alpha\ra^2\la\pi\,\hat\alpha\ra^2}\,, \]
where we have fixed $\beta = \hat\alpha$. Our boundary conditions are $\bar\cA|_{\pi\sim\alpha} = \bar\cA|_{\pi\sim\hat\alpha}=0$.\\
	
To determine the corresponding space-time theory we proceed in the usual way by fixing the gauge using a frame field $\sigha:\PT\to G$. It has the usual redundancies $\sigha\mapsto h\sigha g^{-1}$. The holomorphic Wilson line from $\alpha$ to $\hat\alpha$,
\[ \cW_{\alpha\to\hat\alpha} = \sigma = \sigma_{\hat\alpha}^{-1}\sigma_\alpha\,, \]
exhausts the gauge invariant data that can be extracted from $\sigha$. (For convenience we write $\sigma_\alpha = \sigha|_{\pi\sim\alpha}$ and $\sigma_{\hat\alpha} = \sigha|_{\pi\sim\hat\alpha}$.) Our reality conditions imply that the combination
\[ (C^*\sigha)\Theta(\sigha^{-1}) = \rho \]
is independent of $\pi$, but can in principle depend on $x$. $\rho$ obeys
\[ \rho = C^*\rho = \sigha\Theta(C^*\sigha^{-1}) = (\Theta(C^*\sigha)\sigha^{-1})^{-1} = \Theta(\rho^{-1})\,. \]
It's invariant under the right action on $\sigha$, but is not under the left action:
\[ \rho\mapsto h\rho\Theta(h^{-1})\,. \]
We refer to this as congruency. Under the assumption that $\rho$ lies in the same congruency class as the identity matrix, we may fix $\rho = \id$. This does not completely fix the ambiguity in $\sigha$: we may still act on the left by $h:\E^4\to G_\R$, which may be interpreted as the gauge redundancy of the real ASD connection on space-time. We cannot eliminate this residual symmetry by fixing the value $\sigha$ at any point in $\CP^1_x$, since it need not take values in $G_\R$ anywhere. The holomorphic Wilson line from $\alpha$ to $\hat\alpha$ is
\[ \sigma = \sigma_{\hat\alpha}^{-1}\sigma_\alpha = \Theta(\sigma_\alpha^{-1})\sigma_\alpha\,, \]
and obeys
\[ \sigma = \Theta(\sigma^{-1})\,. \]
In particular $\sigma$ does not take values in a Lie group. For example, if we take $G_\R=\SU_n$, then $\sigma$ is a Hermitian form. This rather strange result for the reality condition on Yang's matrix has been observed elsewhere \cite{yang1977condition,crane1987action}.\\
	
The resulting space-time action is just that of \WZWt{4}, given in equation \eqref{eq:WZW4}, taking $\sigma = \Theta(\sigma_\alpha^{-1})\sigma_\alpha$ as its argument. This is obviously invariant under $\sigma_\alpha\mapsto h\sigma_\alpha$ for $h:\E^4\to G_\R$, and so should be viewed as a $\sigma$-model on the coset space $G_\R\backslash G$. In subsection \ref{subsec:HCStoWZW4} we observed that for $\beta=\hat\alpha$ the self-dual 2-form $\mu_{\alpha,\beta}$ appearing in the action of \WZWt{4} is proportional to the K\"{a}hler form in the complex structure determined by $\alpha$. Our treatment here now justifies this choice.\\
	
The Lax connection is determined by a space-time gauge field $\hat\cA'_A = A_{AA'}\pi^{A'}$, and our boundary conditions imply
\[ A_{AA'} = \alpha_{A'}{\hat\alpha}^{B'}\partial_{AB'}\Theta(\sigma_\alpha)\Theta(\sigma_\alpha^{-1}) - \hat\alpha_{A'}\alpha^{B'}\partial_{AB'}\sigma_\alpha\sigma_\alpha^{-1}\,. \]
It is clear that $A = \theta(A)$, and that the left action $\sigma_\alpha\mapsto h\sigma_\alpha$ really does correspond to a space-time $G_\R$ gauge symmetry.\\

Unfortunately the constraint $\Omega=\overline{C^*\Omega}$ is rather restrictive. Most disappointingly, it precludes us making the choice $\Omega = \D^3Z/(Z\cdot A)^4$ which was used to generate the LMP action in subsection \ref{subsec:LMP}.


\subsubsection*{Ultrahyperbolic signature}

We now turn our attention to the case of ultrahyperbolic, \textit{i.e.}, $(2,2)$ signature, space-time. The reality conditions are essentially identical to those used in the Euclidean case, and are also closely related to the corresponding conditions imposed in \CSt{4} \cite{delduc2019integrable}. 

Writing $C:(x,\pi)\mapsto(x,\bar\pi)$ for spinor conjugation, where $\bar\pi$ is the component-wise complex conjugate of $\pi$, we require that the gauge field $\cA$ on the correspondence space obeys
\[ C^*\cA = \theta(\cA)\,. \label{eq:realityconditionAU} \]
This will be compatible with the gauge redundancies 
\[
\cA\mapsto\cA + \la\diff\pi\,\pi\ra \delta\cA_0 + \diff x^{AA'}\pi_{A'}\delta\cA_A\,,\qquad
\cA\mapsto g\cA g^{-1} - \diff g g^{-1}
\]
provided the gauge parameters likewise obey $C^*\delta\cA_0 = \theta(\delta\cA_0)$, $C^*\delta\cA_A = \theta(\delta\cA_A)$, and $C^*g = \Theta(g)$. The action \eqref{eq:genericCS} on $\cF$ will then be real if in addition
\[ \Omega = \overline{C^*\Omega}\,. \]
This constraint is far easier to satisfy in ultrahyperbolic as compared to Euclidean signature, since spinor conjugation fixes a circle in $\CP^1$. For example, we could simply assume that all poles and zeros of $\Omega$ lie on this real circle. In particular, unlike in Euclidean signature, it is now possible to obtain a real LMP action by taking
\[ \Omega = \frac{\D^3Z}{(Z\cdot A)^4} = \frac{\la\diff\pi\,\pi\ra\wedge\diff^2x^{A'B'}\pi_{A'}\pi_{B'}}{2\la\pi\,\alpha\ra^4}\,, \]
with $\alpha = \bar\alpha$. It is easy to see that the resulting effective space-time action in this case is the LMP action \eqref{eq:SLMP} with $\phi$ taking values in $\fg_\R$. The infinitesimal conformal symmetries \eqref{eq:hiddenLMPdilation} \& \eqref{eq:hiddenLMPspecial} of the LMP Lagrangian can be consistently realised on ultrahyperbolic space-time.\\

It also instructive to reconsider
\[ \Omega= \frac{\D^3Z}{(Z\cdot A)^2(Z\cdot B)^2} = \frac{\la\diff\pi\,\pi\ra\wedge\diff^2x^{A'B'}\pi_{A'}\pi_{B'}}{2\la\pi\,\alpha\ra^2\la\pi\,\beta\ra^2}\,. \]
If $\alpha,\beta$ are real dual twistors then the holomorphic Wilson line from $\alpha$ to $\beta$ takes values in $G_\R$. This descends to \WZWt{4} with values in $G_\R$. Alternatively we can take $\beta = \bar\alpha$, which is similar to the Euclidean case. It leads to \WZWt{4} with argument $\sigma = \Theta(\sigma_\alpha^{-1})\sigma_\alpha$ for some $\sigma_\alpha:\R^4\to G_\R\backslash G$.\\

Finally, in the case where we impose trigonometric boundary conditions at pairs of simple poles in $\Omega$ the appropriate reality conditions are a little more involved. If the two simple poles both occur at real spinors, then we choose $\fg_\pm$ to be Lagrangian subalgebras of $\fg_\R$, so that $(\fg_\R,\fg_-,\fg_+)$ is a real Manin triple. In the case where the simple poles occur at conjugate spinors it is natural to relax our boundary conditions somewhat so that the residues generated at the two poles cancel one another. For further details we refer the reader to \cite{delduc2020unifying}, where analogous boundary conditions were introduced in \CSt{4}.\\

Whilst ultrahyperbolic signature is of less physical interest, it is often more convenient as a starting point for generating 2d integrable theories via symmetry reduction. This is because of the greater flexibility it offers in reductions and reality conditions \cite{mason1996integrability}. In particular, the reality conditions for \CSt{4} introduced in \cite{delduc2020unifying} are recovered by performing reductions in ultrahyperbolic signature.


\section{5d Chern-Simons on minitwistor correspondence space} \label{sec:HCStoCS5}

In this section we consider symmetry reductions by 1 dimensional groups of translations. The most interesting integrable system known to arise in this manner is the Bogomolny equation describing magnetic monopoles. We will find that it can be described by a 5d Chern-Simons (\CSt{5}) theory on minitwistor correspondence space, $\PN$. Indeed, we have the following relationships between CS and integrable theories.
\begin{figure}[h!] \label{fig:cdbogomolny}
	\centering
	\begin{tikzcd}
		& \begin{tabular}{@{}c@{}}Holomorphic Chern-Simons \\ theory on $\PT$ \end{tabular} \arrow[dl,"\text{symmetry reduction}",labels=above left] \arrow[dr,"\text{solving along fibres}"] & \\
		\begin{tabular}{@{}c@{}} 5d Chern-Simons \\ theory on $\PN$ \arrow[dr,"\text{solving along fibres}",labels=below left] \end{tabular}  
		& & \begin{tabular}{@{}c@{}} 4d integrable \\ theory on $\E^4$ \end{tabular} \arrow[dl,"\text{symmetry reduction}",labels=below right] \\ & \begin{tabular}{@{}c@{}} 3d integrable \\ theory on $\E^3$ \end{tabular} & 
	\end{tikzcd}
	\caption{A guide the relationship between CS type theories and integrable systems in dimensions 3 and 4}
\end{figure}\\
\CSt{5} on minitwistor correspondence space is a purely bosonic counterpart of the super minitwistor correspondence space action for the supersymmetric Bogomolny equations introduced in \cite{popov2005topological,adamo2018minitwistors}.


\subsection{Minitwistor correspondence} \label{subsec:minitwistor}

It is well known that the Bogomolny equations arise as a symmetry reduction of the ASDYM equations on 4d Euclidean space-time by a 1 dimensional group of translations. In much the same way as ASD connections on 4d space-time are described by holomorphic vector bundles on twistor space, solutions to the Bogomolny equations are described by holomorphic bundles over a complex manifold known as minitwistor space \cite{ward1981yang,hitchin1982monopoles}. This goes by the name of the Hitchin-Ward correspondence. In this subsection we review how minitwistor space arises as a quotient of $\PT$ by a translation.\\

Consider the quotient of $\E^4$ by the 1 dimensional group of translations $\cH^+$ generated by the real vector $X$. The orbits of $\cH^+$ each intersect the 3-plane $\E^3\cong\{x^{AA'}X_{AA'}=0\}\subset\E^4$ once, allowing us to identify it with the quotient $\cH^+\backslash\E^4$. We write $\iota:\E^3\to\E^4$ for the embedding of this subspace into $\E^4$. It admits natural coordinates
\[ y^{A'B'} = \varepsilon_{AB}x^{AA'}X^{BB'}\,, \]
where $y^{A'B'} = y^{B'A'}$. The choice of vector $X$ breaks the $\SO_4(\R)$ space-time symmetry to $\SO_3(\R)$, allowing us to identify primed and unprimed spinor indices. For further details see appendix \ref{app:Not}. Assuming that $\delta(X,X)=2$ the standard flat metric $\delta(y,y) = y^{A'B'}y_{A'B'}$ is induced on $\E^3$.\\

To perform this quotient on twistor space we must lift $X$ to a vector field on $\PT$ which respects the complex structure. For a translation this lift is trivial: $\cX = X^{AA'}\p_{AA'}$. Let $\PN$ be the quotient of $\PT$ by the translations generated by $\cX$. $\PN$ can be identified with $\iota^*\PT$, the pullback of the fibre bundle $\PT\xrightarrow{\Pi}\E^4$ by the embedding $\iota$. We abuse notation by writing $\iota:\PN\hookrightarrow\PT$. As a smooth manifold $\PN\cong\E^3\times\CP^1$, and we use the coordinates $(y,\pi)$ accordingly. It will be useful to introduce
\[ n^{A'B'} = \frac{i(\pi^{A'}{\hat\pi}^{B'} + \hat\pi^{A'}\pi^{B'})}{\sqrt{2}\|\pi\|^2}\,, \]
a unit vector in $\E^3$ which is smoothly parametrised by $\CP^1$. Pushing forward the holomorphic structure by the quotient map induces a `partially holomorphic' structure on $\PN$. (See \cite{rawnsley1979} for a discussion of partially holomorphic structures.) This is determined by the integrable subbundle locally generated by $\{\bar\p_0,\pi^{B'}\p_{A'B'}\}$. Following \cite{jones1984minitwistors,jones1985minitwistor}, we refer to the partially holomorphic manifold $\PN$ as minitwistor correspondence space.\footnote{$\PN$ is also the space of light rays in four dimensional Minkowski space: the pair $(y,\pi)$ uniquely determines a ray in $\M^{1,3}$ passing through the point $y$ on a constant time slice in the direction $\pi^{A'}\bar\pi^A$. From this perspective, $\PN$ is the real codimension 1 slice $\{Z\cdot\bar Z=0\}\subset\PT$, from which $\PN$ inherits its partially holomorphic structure as a CR manifold.}\\

We can also exploit the complex structure on $\PT$ to take a different quotient. The vector field $\cX$ can be split as $\cX^{(1,0)}+\cX^{(0,1)}$ where
\be \cX^{(1,0)} = -\frac{X^{AA'}{\pi}_{A'}{\hat\pi}^{B'}\p_{AB'}}{\|\pi\|^2}\,,\qquad \cX^{(0,1)} = \frac{X^{AA'}{\hat\pi}_{A'}\pi^{B'}\p_{AB'}}{\|\pi\|^2} \ee
take values in the holomorphic and antiholomorphic tangent bundles respectively. Together these vector fields generate an action of $\C$, the complexification of $\cH^+$. The quotient of $\PT$ by this group is a complex manifold, which we refer to as minitwistor space $\MT$. It can be obtained as the quotient of minitwistor correspondence space by the pushforward of
$i(\chi^{(0,1)}-\chi^{(1,0)})$, which is
\[ \frac{i(\pi^{A'}{\hat\pi}^{B'} + {\hat\pi}^{A'}\pi^{B'})}{\|\pi\|^2}\p_{A'B'} = \sqrt{2} n^{A'B'}\p_{A'B'} \,. \]
This vector field generates translations in the direction $n$, and quotienting by it gives the space of oriented in lines in $\E^3$. We therefore have the minitwistor correspondence \cite{hitchin1982monopoles}.
\[ \begin{tikzcd}
	& \PN \arrow{dl}[above]{P} \arrow{dr}{\Pi} & \\
	\MT & & \E^3
\end{tikzcd} \]
$\gamma = y^{A'B'}\pi_{A'}\pi_{B'}$ is invariant under the translation generated by $n^i\p_i$, and so descends to the quotient. $(\pi,\gamma)$ are then homogeneous holomorphic coordinates on $\MT$, with $(\pi,\gamma)\sim (t\pi,t^2\gamma)$ for $t\in\C^*$. As a complex manifold $\MT\cong T^{1,0}\CP^1\cong(\cO(2)\to\CP^1)$ with $\gamma$ the holomorphic coordinate along the fibres. Explicitly $P:(x,\pi) \mapsto (\gamma,\pi) = (y^{A'B'}\pi_{A'}\pi_{B'},\pi)$.\\

As a partially holomorphic manifold $\PN\cong\MT\times\R$ with partial connection
\[ \diff' = \bar\p_\MT + \diff_\R\,. \]
We will find it convenient to introduce the following frame for the integrable subbundle $\la\{\bar\p_0,\pi^{A'}\p_{A'B'}\}\ra_\C$ determining the partially holomorphic structure on $\PN$:
\be \label{eq:PNvecs} \bar\p_0 = \|\pi\|^2\pi^{A'}\frac{\p}{\p\hat\pi^{A'}}\,,\qquad \bar\p_\gamma = \pi^{A'}\pi^{B'}\p_{A'B'}\,,\qquad \p_t = n^{A'B'}\p_{A'B'} = \frac{\sqrt{2}i\pi^{A'}{\hat\pi}^{B'}\p_{A'B'}}{\|\pi\|^2}\,. \ee
A corresponding set of $1$-forms is
\be \label{eq:PNforms}\bar e^0 = \frac{\la\diff\hat\pi\,\hat\pi\ra}{\|\pi\|^4}\,,\qquad \bar e^\gamma = \frac{\diff y^{A'B'}\hat\pi_{A'}\hat\pi_{B'}}{\|\pi\|^4}\,,\qquad e^t = \diff y^{A'B'} n_{A'B'} = \frac{\sqrt{2}i\diff y^{A'B'}\pi_{A'}{\hat\pi}_{B'}}{\|\pi\|^2}\,. \ee
Together with $\la\diff\pi\,\pi\ra,\diff y^{A'B'}\pi_{A'}\pi_{B'}$ these form a frame for $T^*\PN$. In terms of these frames
\be \diff'= \bar e^0\bar\p_0 + \bar e^\gamma\bar\p_\gamma + e^t\p_t\,. \ee


\subsection{5d Chern-Simons as a symmetry reduction of Holomorphic Chern-Simons}

We can now apply the symmetry reduction by our 1 dimensional group of translations to HCS with measure $\Omega = \Phi\D^3Z$. In the usual way we will not compactify in the $\cX$ direction, but instead simply discard the divergent integral. This is achieved by contracting the vector $\cX$ into the Lagrangian, which saturates the components in the invariant direction, and then pulling back by $\iota:\PN\to\PT$. We, of course, also assume that $\cL_\cX\bar\cA = 0$.\\

We find that
\[ \begin{aligned}
&\iota^*\big(\cX\iprod(\Omega\wedge\HCS(\bar\cA))\big) = \iota^*\big(\cX\iprod(\Omega\wedge\mathrm{CS}(\bar\cA))\big) \\
&= (\Phi\la\diff\pi\,\pi\ra\wedge\diff y^{A'B'}\pi_{A'}\pi_{B'})\wedge\mathrm{CS}(A') = (\Phi\la\diff\pi\,\pi\ra\wedge\diff\gamma)\wedge\PHCS(A')
\end{aligned} \]
where
\be \label{eq:PNA} A' = \iota^*\bar\cA + \frac{i}{\sqrt{2}}\diff y^{A'B'}n_{A'B'}\iota^*(\cX\iprod\bar\cA) \ee
and
\[ \PHCS(A') = \tr\bigg(A'\diff'A' + \frac{2}{3}A'\wedge A'\wedge A'\bigg) \]
for $\diff' = \bar\p_\MT + \diff_\R$. $A'$ can be expanded in terms of the 1-forms introduced in equation \eqref{eq:PNforms}:
\[ A' = \bar e^0\bar A_0' + \bar e^\gamma\bar A_\gamma' + e^tA_t'\,. \]
From equation \eqref{eq:PNA} we can see that $A'$ inherits all of its boundary conditions from those on $\bar\cA$. In particular the reduction does not generate poles, as it did in subsection \ref{subsec:HCStoCS4}. Introducing
\[ \omega = \Phi\la\diff\pi\,\pi\ra\wedge\diff\gamma\,, \]
the pullback by $P:\PN\to\MT$ of a meromorphic from on $\MT$, we can write the action on the quotient as
\be \label{eq:CS5} S_{\CS{5}}[A'] = \frac{1}{2\pi i}\int_\PN\omega\wedge\PHCS(A')\,. \ee
This is as partially holomorphic 5d Chern-Simons theory on $\PN$, which we abbreviate as \CSt{5}.\\

We now make explicit the connection between this theory and the Bogomolny equations on $\E^3$. In order to do so we proceed in the usual way by trivialising $\bar A_0'$ using a frame field $\sigha:\PN\to G$. Up to gauge this can be expressed in terms of a set of group and Lie algebra fields on $\E^3$ determined by the boundary conditions. Performing a forbidden gauge transformation by $\sigha$ brings us into the gauge $\bar A_0' = 0$ in which the classical equations
\[ F'(A') = \diff'A' + [A',A'] = 0 \]
simplify. Even in this gauge the curvature must be computed with some care, since
\[ \diff'\bar e^0 = 0\,,\qquad\diff'\bar e^\gamma = \frac{2{\bar e}^0\wedge\diff y^{A'B'}\pi_{A'}{\hat\pi}_{B'}}{\|\pi\|^2} = -\sqrt{2}i{\bar e}^0\wedge e^t\,,\qquad\diff'e^t = 0\,. \]
We find that
\be \label{eq:CS5eom} F'(A') = \bar e^0\wedge\bar e^\gamma\bar\p_0\bar A_\gamma + \bar e^0\wedge e^t(\bar\p_0A_t - \sqrt{2}i\bar A_\gamma) + \bar e^\gamma\wedge e^t (\bar\p_\gamma A_t - \p_t\bar A_\gamma + [\bar A_\gamma,A_t]) = 0\,. \ee
From the components involving $\bar e^0$ we have
\[ \bar\p_0\bar A_\gamma' = 0\,,\qquad \bar\p_0A_t' - \sqrt{2}i\bar A_\gamma' = 0\,. \]
Assuming that $A'$ is free from poles, the first of these is simply the statement that $\bar A_\gamma$ is holomorphic in $\pi$. Since $\bar A_\gamma'$ has holomorphic weight 2
\[ \bar A_\gamma' = \pi^{A'}\pi^{B'}a_{A'B'} \]
where $a_{A'B'} = a_{B'A'}$ depends only on $y$. Substituting this into the second equation gives
\[ \bar\p_0A_t' = \sqrt{2}i\pi^{A'}\pi^{B'}a_{A'B'}\,, \]
the solution to which is
\[ A_t' = \frac{\sqrt{2}i\pi^{A'}{\hat\pi}^{B'}a_{A'B'}}{\|\pi\|^2} + i\varphi = n^{A'B'}a_{A'B'} + i\varphi \,. \]
Here $\varphi$ also depends only on $y$. In full we therefore have
\[ A' = {\bar e}^\gamma\pi^{A'}\pi^{B'}a_{A'B'} + e^t(n^{A'B'}a_{A'B'} + i\varphi)\,. \]
Recall that there is a redundancy in the choice of frame field $\sigha\mapsto h\sigha$ for $h$ independent of $\pi$. Under this transformation $a$ and $\varphi$ transform as a connection and adjoint valued scalar on $\E^3$. The boundary conditions on $A'$ allow us to express $a$ and $\varphi$ in terms of the group and Lie algebra fields determining $\sigha$. At this point the $\pi$ dependence of $A$ is in principle completely fixed, and integrating over the fibres of $\PN\xrightarrow{\Pi}\E^3$ gives a 3d space-time action.\\

Its classical equations of motion will imply that the remaining component in \eqref{eq:CS5eom} vanishes
\be \label{eq:Bogomolnyalg} \bar\p_\gamma A_t' - \p_t\bar A_\gamma' + [\bar A_\gamma',A_t'] = \frac{{\sqrt 2}i\pi^{A'}\pi^{B'}\pi^{C'}{\hat\pi}^{D'}}{\|\pi\|^2}\bigg(f_{A'B'C'D'}(a) + \frac{1}{\sqrt{2}}\varepsilon_{C'D'}\nabla_{A'B'}\varphi\bigg) = 0\,, \ee
where $\nabla = \diff + a$ and $f(a)$ is the curvature. Decomposing $\pi^{A'}\pi^{B'}\pi^{C'}{\hat\pi}^{D'}$ into its totally symmetric and mixed parts
\[ 4\pi^{A'}\pi^{B'}\pi^{C'}{\hat\pi}^{D'} = 4\pi^{(A'}\pi^{B'}\pi^{C'}\hat{\pi}^{D')} - \|\pi\|^2(\pi^{A'}\pi^{B'}\varepsilon^{C'D'} + \pi^{A'}\pi^{C'}\varepsilon^{B'D'} + \pi^{B'}\pi^{C'}\varepsilon^{A'D'}) \]
equation \eqref{eq:Bogomolnyalg} is
\[ \pi^{A'}\pi^{B'}\varepsilon^{C'D'}f_{A'C'B'D'} = \frac{1}{\sqrt{2}}\pi^{A'}\pi^{B'}\nabla_{A'B'}\varphi\,. \]
We then identify
\[ (\ast f)_{A'B'} = \sqrt{2}f_{A'C'B'D}\varepsilon^{C'D'}\,, \]
and so the Bogomolny equation follows
\[ \ast f = \nabla\varphi\,. \]
This is essentially the statement of the Hitchin-Ward correspondence.\\

So far we have not been specific about our choice of $\Phi$ and associated boundary conditions, except in assuming that we are not permitting any poles in $A'$. By choosing $\Phi$ as in sections \ref{sec:sec2} and \ref{sec:sec3} we obtain a range of space-time actions which are all straightforward to compute as symmetry reductions of those appearing therein. If $\Phi$ is nowhere vanishing then we need not allow poles in $A'$, and so the classical equations of motion of the resulting 3d theory will be equivalent to the Bogomolny equations. It is also straightforward to generalise the reality conditions in section \ref{sec:ReSt} to the 5d case, though they remain rather stringent.\\

We could also perform an analogous reduction by a non-null translation in ultrahyperbolic signature. This leads to \CSt{5} on the projective spinor bundle of 3d Minkowski space $\M^{2,1}$. Under this reduction the partial connection acquires a simple pole along the locus $\pi = \bar\pi$. In Lorentzian signature the reality conditions on $\Phi$ are less stringent. $\Phi = 1/\la\pi\,\alpha\ra^2\la\pi\,\beta\ra^2$ leads to the $2+1$ dimensional chiral model \cite{ward1988soliton} if $\alpha,\beta$ are real, and \cite{manakov1981three} if they are conjugate. $\Phi = 1/\la\pi\,\alpha\ra^4$ for $\alpha$ real leads to the pseudodual of the 2+1 dimensional chiral model \cite{dimakis2008dispersionless}.


\section{Conclusions}
We have seen that CS type theories on twistor space, and more generally on twistor correspondence spaces, are classically equivalent to 4d integrable theories on space-time. When the $3$-form $\Omega$ is nowhere vanishing these integrable theories have equations of motion equivalent to the ASDYM equations.\\

Furthermore, symmetry reductions of these 4d theories by subgroups of translations give actions for lower dimensional classical integrable field theories. Lifting the action of these translations to twistor space and performing the symmetry reductions there leads to CS type theories on reduced twistor correspondence spaces. In this way we recover $\CS{4}$ descriptions of a range of 2d classical integrable field theories.\\

It is well known that the non-linear Schr\"{o}dinger equation, Korteweg-de Vries equation and Toda field theory can also be obtained as reductions of the ASDYM equations~\cite{mason1996integrability}. It is therefore natural to expect that applying these reductions to CS type theories on twistor correspondence spaces would allow these integrable systems to be realised in \CSt{4}. It would be interesting to see how these reductions are related to the realisations of KdV and Toda theory appearing in \cite{gaiotto2020kondo} and \cite{ashwinkumar20204d} respectively. We leave exploring this possibility for future work.\\

We have also seen that when the $3$-form $\Omega$ has zeros the corresponding 4d space-time theory does not have equations of motion equivalent to the ASDYM equations. It would be interesting to characterise the classical integrable field theories which arise as symmetry reductions for such $\Omega$.\\

Finally, the major advantage of the perspective presented here over the familiar story realising integrable systems as symmetry reductions of the ASDYM equations is that it's performed at the level of the action. This is the first step towards a quantum treatment.

\newpage

\appendix


\section{Notation and Background}


\subsection{Notation and conventions for spinors} \label{app:Not}

All indices will be regarded as `abstract' in the sense that $V^a$ refers to a particular vector, not its components in some basis.\\

Roman indices $a,b,c,\dots$ from the beginning of the alphabet label elements of the tangent (and cotangent) bundles to 4d complexified Minkowski space, $\CM^4$. Fixing a real structure on $\CM^4$ these become labels for elements of the tangent (and cotangent) bundles to a real form of $\CM^4$, {\it e.g.}, 4d Euclidean space $\E^4$. They are contracted using the ${\rm SO}_4(\C)$-invariant tensor $g_{ab}$. We also make use of the invariant alternating tensor $\varepsilon_{abcd}$ with $\varepsilon_{0123}=1$. Primed and unprimed capital indices $A',B',C',\dots$ and $A,B,C,\dots$ label elements of $\bbS^+$ and $\bbS^-$, the left and right handed spin bundles over 4d complexified Minkowski space $\CM^4$ respectively. Fixing a real structure they can be similarly interpreted as labels for the spin bundles over a real form of $\CM^4$. They are contracted using the ${\rm SL}_2(\C)$-invariant tensors $\varepsilon_{A'B'}$ and $\varepsilon_{AB}$ where $\varepsilon_{0'1'} = \varepsilon_{01} = 1$. We often write
\[ \la\alpha\,\beta\ra = \alpha^{A'}\beta^{B'}\varepsilon_{A'B'} = \alpha^{A'}\beta_{A'}\,\quad [\mu\,\nu] = \mu^A\nu^B\varepsilon_{AB} = \mu^A\nu_A\,. \]
The isomorphism $\CM^4\cong\bbS^+\otimes\bbS^-$ allows us to identify $V^a = V^{AA'}$. Then
\[ g_{ab} = \varepsilon_{AB}\varepsilon_{A'B'}\,,\quad \varepsilon_{abcd} = \varepsilon_{AC}\varepsilon_{BD}\varepsilon_{A'D'}\varepsilon_{B'C'} - \varepsilon_{AD}\varepsilon_{BC}\varepsilon_{A'C'}\varepsilon_{B'D'}\,. \]
We use Roman indices $i,j,k,\dots$ from the middle of the alphabet to label elements of the tangent (and cotangent) bundles to 3d complexified Minkowski space, $\CM^3$. We can also view these as indices on real forms of $\CM^3$, {\it e.g.}, 3d Euclidean space $\E^3$. They are contracted using the ${\rm SO}_3(\C)$-invariant tensor $g_{ij}$, and we also make use of the invariant alternating tensor $\varepsilon_{ijk}$ with $\varepsilon_{123}=1$. We abuse notation by using unprimed spinor indices $A',B',C',\dots$ as labels for elements of $\bbS$, the spin bundle over $\CM^3$. These indices are contracted using $\varepsilon_{A'B'}$ as above. The isomorphism $\CM^3\cong S^2\bbS$ allows us to identify $V^i = V^{A'B'}$ where the right hand side is symmetric under exchange of $A'$ and $B'$. Then
\[ g_{ij} = \frac{1}{2}(\varepsilon_{A'C'}\varepsilon_{B'D'} + \varepsilon_{A'D'}\varepsilon_{B'C'})\,,\quad \varepsilon_{ijk} = \frac{1}{\sqrt{2}}(\varepsilon_{A'C'}\varepsilon_{E'B'}\varepsilon_{D'F'} - \varepsilon_{A'F'}\varepsilon_{C'E'}\varepsilon_{B'D'})\,, \]
where we are identifying $i=A'B'$, $j=C'D'$ and $k=E'F'$. Our abuse of notation is justified by the observation that $\bbS$ can be viewed as the pullback of $\bbS^+$ by
\[ \iota:\CM^3\xhookrightarrow{}\CM^4\,,\quad x^{A'B'}\mapsto x^{AA'}=\sqrt{2}\varepsilon_{B'C'}X^{AB'}x^{AC'} \]
for $X$ some choice of unit vector in $\CM^3$.\\

We use Greek indices $\alpha,\beta,\gamma,\dots$ to label elements of $\C^4$, {\it i.e.}, for twistor indices.


\subsection{Homogeneous coordinates on $\CP^1$} \label{app:homcoord}
	
Throughout the paper we use homogeneous coordinates on $\CP^1$, which often appears as the fibre of the left-handed projective spinor bundle over space-time. Writing $\pi^{A'} = (\pi^{0'},\pi^{1'}) \in \C^2\setminus\{0\}$ for a non-zero left-handed spinor, we represent its equivalence class under the relation $\pi^{A'}\sim t\pi^{A'}$ for $t\in\C^*$ by $[\pi]$. A function $F$ of $\pi^{A'}$ and its complex conjugates is said to have holomorphic weight $m$ and antiholomorphic weight $n$ if under this rescaling $F\mapsto t^m\bar t^n F$. We can interpret $F$ as a smooth section of the line bundle $\cO(m)\otimes\bar\cO(n)$.\\

Rescalings of $\pi$ are generated by the vector field $\Gamma = \pi^{A'}\p_{\pi^{A'}}$ and its conjugate $\bar\Gamma$. We can identify $T^{1,0}_{[\pi]}\CP^1$ with the quotient of $T^{1,0}_\pi\C^2$ by the subspace generated by $\Gamma$. These subspaces generate a subbundle $\langle\Gamma\rangle\subset T^{1,0}\C^2$. Sections of $T^{1,0}\CP^1\otimes\cO(m)\otimes\bar\cO(n)$ are then realised as sections of the quotient bundle $T^{1,0}\C^2/\langle\Gamma\rangle$ with holomorphic weight $m$ and antiholomorphic weight $n$. The line bundle $T^{1,0}\CP^1\otimes\cO(-2)$ has a unique holomorphic section given by
\be \p_0 = \bigg[-\frac{\alpha^{A'}}{\la\pi\,\alpha\ra}\frac{\p}{\p\pi^{A'}}\bigg]\,, \ee
for $[\alpha]\in\CP^1$ an arbitrary choice of reference spinor. The above definition makes sense only for $[\pi]\neq[\alpha]$, but the equivalence class on the right hand side is actually independent of $\alpha$ and so can be used to define $\p_0$ globally. This whole discussion goes through identically for the antiholomorphic tangent bundle if we replace $\Gamma$ by $\bar\Gamma$ everywhere.\\

The holomorphic cotangent space $T^{*\,1,0}_{[\pi]}\CP^1$ can be realised as the kernel of $\iota_\Gamma$ in $T^{*\,1,0}_\pi\C^2$. This defines a subbundle $\ker\iota_\Gamma\subset T^{*\,1,0}\C^2$. Sections of $T^{*\,1,0}\CP^1\otimes\cO(m)\otimes\bar\cO(n)$ can be identified with sections of the subbundle $\ker\iota_\Gamma$ with holomorphic and antiholomorphic weights $m$ and $n$ respectively. The line bundle $T^{*\,1,0}\CP^1\otimes\cO(2)$ has a unique holomorphic section
\be e^0 = \la\diff\pi\,\pi\ra\,. \ee
The antiholomorphic cotangent space and higher degree forms can be incorporated into this description in the obvious way.\\

Fixing a dyad of left-handed spinors $\{\alpha,\beta\}$ we may introduce inhomogeneous coordinates on $\CP^1$ by
\[ \pi^{A'} \sim \alpha^{A'} - \zeta\beta^{A'}
\quad\text{where}\quad\zeta = \frac{\la\pi\,\alpha\ra}{\la\pi\,\beta\ra}\,. \]
Then
\be \diff\zeta = \frac{e^0}{\la\pi\,\beta\ra^2}\,,\qquad \frac{\p}{\p\zeta} = \la\pi\,\beta\ra^2\p_0\,. \ee
Using these identities we can easily transition from homogeneous to inhomogeneous coordinates.


\subsection{The anti-self-dual Yang-Mills equations} \label{app:ASDYM}
The ASDYM equations for a connection $\nabla = \diff + A$ on a principal $G$-bundle over $\CM^4$ are given by
\[ \ast_4 F = - F\,, \]
for $\ast_4$ the Hodge star operator induced by the metric $g$ and $F$ the curvature of $\nabla$. The ASDYM equations imply the Yang-Mills equations as a consequence of the Bianchi identity on $F$,
\[ \nabla\ast_4F = - \nabla F = 0 \]
Since the curvature is an anti-symmetric tensor it can be decomposed as
\[ F_{ab} = \varepsilon_{AB}F_{A'B'} + \varepsilon_{A'B'}F_{AB} \]
for $F_{AB}$ and $F_{A'B'}$ symmetric in their indices. The first term in this sum is SD, and the second is ASD, and so the ASDYM equations can be expressed as
\[ F_{A'B'} = - \frac{1}{2}\varepsilon^{AB}[\nabla_{AA'},\nabla_{BB'}] = 0\,. \]
Alternatively they are equivalent to the vanishing of
\[ \pi^{A'}\pi^{B'}[\nabla_{AA'},\nabla_{BB'}] \]
for all $[\pi]\in\CP^1$. This is the statement that the restriction of $\nabla$ to any SD 2-plane in $\CM^4$ is flat. It's this observation that allowed Ward to relate solutions of the ASDYM equations to holomorphic vector bundles over the twistor space of $\CM^4$ \cite{ward1977self}. There are two 2\textsuperscript{nd} order forms for the ASDYM equations frequently discussed in the literature. Both require breaking Lorentz invariance in some form.\\

To obtain the first we fix a dyad $\la\alpha\,\beta\ra = 1$. We then solve
\[ \alpha^{A'}\alpha^{B'}[\nabla_{AA'},\nabla_{BB'}] = 0\,, \]
by writing $\alpha^{A'}A_{AA'} = -\alpha^{A'}\p_{AA'}\sigma_\alpha\sigma^{-1}_\alpha$ for some $\sigma_\alpha:\CM^4\to G$, and similarly solve
\[ \beta^{A'}\beta^{B'}[\nabla_{AA'},\nabla_{BB'}] = 0\,, \]
by writing $\beta^{A'}A_{AA'} = -\beta^{A'}\p_{AA'}\sigma_\beta\sigma^{-1}_\beta$ for some $\sigma_\beta:\CM^4\to G$. A gauge transformation by $\sigma_\beta^{-1}$ fixes $\beta^{A'}A_{AA'}=0$, and we find that
\[ A_{AA'} = -\beta_{A'}\alpha^{B'}\p_{AB'}\sigma\sigma^{-1} \]
for $\sigma = \sigma_\beta^{-1}\sigma_\alpha$. $A$ is then ASD if
\[ \varepsilon^{AB}\alpha^{A'}\beta^{B'}[\nabla_{AA'},\nabla_{BB'}] = \varepsilon^{AB}\alpha^{A'}\beta^{B'}\p_{BB'}(\p_{AA'}\sigma\sigma^{-1}) = 0\,. \]
This equation is known as Yang's equation, and $\sigma$ is referred to as the Yang matrix~\cite{yang1977condition}.\\

For the second we choose a left-handed spinor $\alpha$, which we use to specify the gauge $\alpha^{A'}A_{AA'} = 0$. This is solved by $A_{AA'} = \alpha_{A'}\xi_A$ for some right-handed spinor field $\xi$. Next we impose
\[  \varepsilon^{AB}\alpha^{A'}[\nabla_{AA'},\nabla_{BB'}] = \alpha_{B'}\varepsilon^{AB}\alpha^{A'}\p_{AA'}\xi_B = 0 \implies \alpha^{A'}\p_{AA'}\xi^A = 0 \,. \]
From the Poincar\'{e} Lemma we can then write $\xi_A = \alpha^{A'}\p_{AA'}\phi$ for $\phi$ a $\fg$-valued scalar field. The ASDYM equations then follow from
\[ \varepsilon^{AB}[\nabla_{AA'},\nabla_{BB'}] = \alpha_{B'}(2\varepsilon^{CD}\alpha^{D'}\p_{CA'}\p_{DD'}\phi + \alpha_{A'}\varepsilon^{CD}\alpha^{C'}\alpha^{D'}[\p_{CC'}\phi,\p_{DD'}\phi]) = 0\,. \]
We therefore learn that
\[ \varepsilon^{CD}\alpha^{D'}\p_{CA'}\p_{DD'}\phi + \alpha_{A'}\varepsilon^{CD}\alpha^{C'}\alpha^{D'}[\p_{CC'}\phi,\p_{DD'}\phi] = 0\,. \]
Contracting with an arbitrary left-handed spinor $\beta$ which is not proportional to $\alpha$ this can be rewritten as
\[ \Delta\phi = \varepsilon^{AB}\alpha^{A'}\alpha^{B'}[\p_{AA'}\phi,\p_{BB'}\phi]\,, \]
where $\Delta = g^{ab}\p_a\p_b$ is the Laplacian.


\subsection{The twistor correspondence in Lorentzian and ultrahyperbolic signature} \label{app:MUtwistors}

The twistor space of complexified space-time, $\C\M^4$, is $\PT = {\CP}^3\setminus{\CP}^1$. In particular each point $(\omega^A,\pi_{A'})\in\PT$ defines a totally null 2-plane in $\C\M^4$ with SD tangent bivector by
\be \omega^A = x^{AA'}\pi_{A'}\,. \label{eq:increl} \ee
We refer to such 2-planes as $\alpha$-planes, and their tangent bivectors are proportional to $\varepsilon^{AB}\pi^{A'}\pi^{B'}$. Conversely fixing $x\in\C\M^4$ in the incidence relation \eqref{eq:increl} and letting $Z^\alpha = (\omega^A,\pi_{A'})$ vary defines the holomorphic line ${\CP}_x\xhookrightarrow{\iota_x}\PT$. We therefore have a double fibration of a correspondence space: the set of pairs $(x,Z)\in\C\M^4\times\PT$ such that equation \eqref{eq:increl} holds.\\

We can view this correspondence space as an open subset of the flag manifold ${\mathbb F}_{(1,2)}{\mathbb C}^4$. ${\mathbb F}_{(1,2)}{\mathbb C}^4$ is the set of vector subspaces $E_1\subset E_2\subset \C^4$ such that $\dim E_1 = 1$ and $\dim E_2 = 2$. We may interpret $E_1$ as a point in $\CP^3$, and $E_2$ as a subspace $\CP^1\subset\CP^3$ containing $E_1$. Under the additional requirement that both $E_1$ and $E_2$ are contained in $\PT\subset\CP^3$ we can identify $E_1$ with a point in $\PT$, and $E_2$ with $\CP_x$ for a unique $x\in\C\M^4$. The condition $E_1\subset E_2$ is equivalent to the incidence relation \eqref{eq:increl}. We therefore identify the open subset of ${\mathbb F}_{(1,2)}{\mathbb C}^4$ defined by $E_1\subset E_2\subset\PT$ with the correspondence space, and we denote it by $\mathbb{F}_{(1,2)}\T$. It is double fibred over $\PT$ and $\C\M$.
\[ \begin{tikzcd}
	& \mathbb{F}_{(1,2)}\T \arrow{dl} \arrow{dr} & \\
	\PT & & \C\M
\end{tikzcd} \]
This is the twistor correspondence for $\C\M^4$. Given a subspace ${\mathcal M}\subset\C\M^4$ we can define a corresponding twistor space $\mathcal{PT} = \{Z\in\PT|Z\cap{\mathcal M}\neq\phi\}$, and have an associated twistor correspondence.
\[ \begin{tikzcd}
	& \mathcal{F} \arrow{dl}[above]{\rho} \arrow{dr}{\pi} & \\
	\mathcal{PT}& & {\mathcal M}
\end{tikzcd} \]
Here ${\mathcal F}$ is the set of pairs $(x,Z)\in{\mathcal U}\times\mathcal{PT}$ obeying the incidence relation \eqref{eq:increl}. ${\mathcal F}$ can naturally be viewed as a subset of ${\mathbb F}_{(1,2)}\T$.\\

Euclidean, Minkowski, and ultrahyperbolic 4-dimensional space-times can be realised as real forms of $\CM^4$, i.e. as the fixed points of an antiholomorphic involution $\CM^4\to\CM^4$. On a real slice the complexified Lorentz group, ${\rm SO}_4(\C)$, reduces to a real form, as does the spin group ${\rm SL}_2(\C)\times {\rm SL}_2(\C)$. A real form can be encoded in a complex conjugation on spinors, which can then be extended naturally to twistors.
\begin{itemize}
	\item Euclidean space $\E^4$: Spinor complex conjugation is given by $\omega^A\mapsto {\hat\omega}^A = (-\overline{\omega^1},\overline{\omega^0})$, $\pi^{A'}\mapsto {\hat\pi}^{A'} = (-\overline{\pi^{1'}},\overline{\pi^{0'}})$. It preserves the handedness of spinors. The extension to twistor space is $\PT\to\PT$, $Z^\alpha = (\omega^A,\pi_{A'})\mapsto {\hat Z}^\alpha = ({\hat\omega}^A,{\hat\pi}_{A'})$.
	\item Minkowski space $\M^4$: Spinor complex conjugation is given by $\omega^A\mapsto {\bar\omega}^{A'} = (\overline{\omega^0},\overline{\omega^1})$, $\pi^{A'}\mapsto {\bar\pi}^A = (\overline{\pi^{0'}},\overline{\pi^{1'}})$. It exchanges the handedness of spinors. The extension to twistor space it is given by $\PT\to \PT^*$, $Z^\alpha = (\omega^A,\pi_{A'})\mapsto {\bar Z}_\alpha = ({\bar\pi}_A,-{\bar\omega}^{A'})$. Note that twistors are mapped to dual twistors.
	\item Ultrahyperbolic space $\bbU^4$: Spinor  complex conjugation is given by $\omega^A\mapsto {\bar\omega}^A = (\overline{\omega^0},\overline{\omega^1})$, $\pi^{A'}\mapsto {\bar\pi}^{A'} = (\overline{\pi^{1'}},\overline{\pi^{0'}})$. It preserves left-handed and right-handed spinors. The extension to twistor space is $\PT\to\PT$, $Z^\alpha = (\omega^A,\pi_{A'})\mapsto {\bar Z}^\alpha = ({\bar\omega}^A,{\bar\pi}_{A'})$.
\end{itemize}
As subspaces of $\CM^4$ we can construct twistor spaces and correspondences for each of the 3 real forms.
\begin{itemize}
	\item Euclidean space $\E^4$: As was discussed in subsection \ref{subsec:TwiE}, the twistor correspondence simplifies dramatically in Euclidean signature. We find that $\mathcal{PT} = \PT$, and the $\alpha$-plane defined by $Z\in\PT$ contains a unique $x\in\E^4$. This $x$ is characterised by the fact that ${\CP}_x^1$ is the projective line connecting $Z$ and ${\hat Z}$. $\PT$ is diffeomorphic to ${\mathcal F}$, which can be identified with $\PS^+$, the left-handed projective spinor bundle over $\E^4$.
	\item Minkowski space $\M^4$: $\mathcal{PT} = \PN = \{(\omega^A,\pi_{A'})\in\PT|Z^\alpha{\hat Z}_\alpha=\omega^A{\bar\pi}_A-\pi_{A'}{\bar\omega}^{A'}=0\}$. A point in $\PN$ defines a light ray in $\M^4$. We can identify ${\mathcal F}$ with $\PS^+$, the left-handed projective spinor bundle over $\M^4$.
	\item Ultrahyperbolic space $\bbU^4$: Following the construction outlined above we have $\mathcal{PT}=\PT$. The fact that complex conjugation on twistors has fixed points means that $\mathcal{PT}=\PT$ has a distinguished real subspace given by $\mathcal{PT}_\R = \PT\cap\PR^3$. Points in $\cT_\R$ correspond to $\alpha$-planes which lie wholly within $\bbU^4$, and with real tangent bivectors $\varepsilon^{AB}\pi^{A'}\pi^{B'}$ for ${\bar\pi} = \pi$. The complement of $\mathcal{PT}_\R$ in $\mathcal{PT}=\PT$ fibres over $\bbU^4$. In particular given $Z\in\mathcal{PT}\setminus\mathcal{PT}_\R$ the projective line connecting $Z$ and ${\bar Z}$ is ${\CP}_x^1$ for some $x\in\bbU^4$. As in the previous two cases we can identify ${\mathcal F}$ with $\PS^+$, the left-handed projective spinor bundle over $\bbU^4$. There is, however, a distinguished real subspace ${\mathcal F}_\R\subset {\mathcal F}$ which is fixed by spinor conjugation. It is an $S^1$ bundle over $\bbU^4$.
\end{itemize}
Note that conjugation on spinors extends by linearity to an antiholomorphic involution on $\CM^4$. This is induces the real form that we started with.


\section{Computations}


\subsection{Calculations for trigonometric actions} \label{app:trigASDYM}

In this appendix we begin by showing that the equations of motion
\[ \big[\mu_{\alpha_-,\alpha_+}\wedge\partial(\sigma_-^{-1}{\tilde\partial}\sigma_+\sigma_+^{-1}\sigma_-)\big]_{\fl_+} = 0\,,\quad \big[\mu_{\alpha_-,\alpha_+}\wedge{\tilde\partial}(\sigma_+^{-1}{\partial}\sigma_-\sigma_-^{-1}\sigma_+)\big]_{\fl_-} = 0
\]
are equivalent to Yang's equation, and therefore to the ASDYM equations.\\

Recall Yang's equation for a $G$-valued field $\sigma$
\[ \mu_{\alpha_-,\alpha_+}\wedge\partial(\tilde\partial\sigma\sigma^{-1}) = 0\iff \mu_{\alpha_-,\alpha_+}\wedge{\tilde\p}(\sigma^{-1}\partial\sigma) = 0\,.
\]
Using our decomposition $U\times{\widetilde H} = L_-L_+$ we can write $\sigma = \sigma_-^{-1}\sigma_+$ for $\sigma_-\in L_-$, $\sigma_+\in L_+$. Substituting this into the two forms of Yang's equation we get
\[ \mu_{\alpha_-,\alpha_+}\wedge\partial(\sigma_-^{-1}{\tilde\p}\sigma_+\sigma_+^{-1}\sigma_- - \sigma_-^{-1}{\tilde\p}\sigma_-) = 0\,,\quad\mu_{\alpha_-,\alpha_+}\wedge{\tilde\partial}(\sigma_+^{-1}\partial\sigma_+ - \sigma_+^{-1}\partial\sigma_-\sigma_-^{-1}\sigma_+) = 0\,. \]
Noting that $\sigma_\pm^{-1}{\tilde\partial}\sigma_\pm$ takes values in $\fl_\pm$, projecting the above equations onto $\fl_+$ and $\fl_-$ respectively gives 
\[ \big[\mu_{\alpha_-,\alpha_+}\wedge\partial(\sigma_-^{-1}{\tilde\partial}\sigma_+\sigma_+^{-1}\sigma_-)\big]_{\fl_+} = 0\,,\quad \big[\mu_{\alpha_-,\alpha_+}\wedge{\tilde\partial}(\sigma_+^{-1}{\partial}\sigma_-\sigma_-^{-1}\sigma_+)\big]_{\fl_-} = 0\,.
\]
These are precisely the classical equations of motion for our action.\\

Showing the converse is marginally less straightforward. We begin by removing the projections in our equations of motion by writing
\be \label{eq:trigeoms} \mu_{\alpha_-,\alpha_+}\wedge(\partial(\sigma_-^{-1}{\tilde\partial}\sigma_+\sigma_+^{-1}\sigma_-) + \rho_-) = 0\,,\quad \mu_{\alpha_-,\alpha_+}\wedge({\tilde\partial}(\sigma_+^{-1}{\partial}\sigma_-\sigma_-^{-1}\sigma_+) + \rho_+) = 0\,,
\ee
for $\rho_\pm\in\fl_\pm$. Conjugating these formulae by $\sigma_-$ and $\sigma_+$ respectively
\[ \begin{aligned} \mu_{\alpha_-,\alpha_+}\wedge([{\tilde\partial}\sigma_+\sigma_+^{-1}\wedge\partial\sigma_-\sigma_-^{-1}] + \partial({\tilde\partial}\sigma_+\sigma_+^{-1}) + \sigma_-\rho_-\sigma_-^{-1})&=0\,, \\
\mu_{\alpha_-,\alpha_+}\wedge([\partial\sigma_-\sigma_-^{-1}\wedge\partial\sigma_+\sigma_+^{-1}] + {\tilde\partial}(\partial\sigma_-\sigma_-^{-1}) + \sigma_+\rho_+\sigma_+^{-1})&=0\,.
\end{aligned} \]
Adding these together gives
\[\mu_{\alpha_-,\alpha_+}\wedge(\partial({\tilde\partial}\sigma_+\sigma_+^{-1}) + \sigma_-\rho_-\sigma_-^{-1} + {\tilde\partial}(\partial\sigma_-\sigma_-^{-1}) + \sigma_+\rho_+\sigma_+^{-1})=0\,, \]
and projecting onto $\fl_\mp$ we find
\[ \mu_{\alpha_-,\alpha_+}\wedge({\tilde\partial}(\partial\sigma_-\sigma_-^{-1}) + \sigma_-\rho_-\sigma_-^{-1}) = 0\,,\quad\mu_{\alpha_-,\alpha_+}\wedge(\partial({\tilde\partial}\sigma_+\sigma_+^{-1}) + \sigma_+\rho_+\sigma_+^{-1}) = 0\,. \]
We can clearly solve these equations to get
\[ \mu_{\alpha_-,\alpha_+}\wedge\rho_- = - \mu_{\alpha_-,\alpha_+}\wedge\partial(\sigma_-^{-1}{\tilde\partial\sigma_-})\,,\quad\mu_{\alpha_-,\alpha_+}\wedge\rho_+ = - \mu_{\alpha_-,\alpha_+}\wedge{\tilde\partial}(\sigma_+^{-1}\partial\sigma_+)\,. \]
Substituting these expressions back into \eqref{eq:trigeoms}, we recover both forms of Yang's equation.\\

Next we show that by substituting
\[ \sigma_-^{-1} = (\ell h,h^{-1})\,,\quad \sigma_+ = (hu,h)\,,\]
into the action
\[ \frac{1}{\la\alpha_-\,\alpha_+\ra}\int_{\E^4}\mu_{\alpha_-,\alpha_+}\wedge\tr(\partial\sigma_-\sigma_-^{-1}\wedge\tilde\partial\sigma_+\sigma_+^{-1})\,,
\]
we obtain \eqref{eq:SMasSpa}.
\[ \begin{aligned}
&\phantom{{}={}} \tr(\p\sigma_-\sigma_-^{-1}\wedge{\tilde\p}\sigma_+\sigma_+^{-1}) \\
&= \tr_0(\partial(h^{-1}\ell^{-1})\ell h\wedge{\tilde\partial}(hu)u^{-1}h^{-1}) - \tr_0(\partial h h^{-1}\wedge{\tilde\partial}h h^{-1}) \\
&= \tr_0(\partial(h^{-1}\ell^{-1})\ell h\wedge h{\tilde\partial}uu^{-1}h^{-1}) + \tr_0(\partial(h^{-1}\ell^{-1})\ell h\wedge \tilde\partial hh^{-1}) - \tr_0(\partial h h^{-1}\wedge{\tilde\partial}h h^{-1}) \\
&= \tr_0(\partial(h^{-2}\ell^{-1})\ell h^2\wedge {\tilde\partial}uu^{-1}) - \tr_0(h^{-1}\partial h\wedge{\tilde\partial}u u^{-1}) - \tr_0(\ell^{-1}\partial\ell\wedge h{\tilde\partial}hh^{-2}) \\
&- \tr_0(h^{-1}\partial h\wedge{\tilde\partial}h h^{-1}) - \tr_0(\partial h h^{-1}\wedge{\tilde\partial}h h^{-1})\,.
\end{aligned} \]
Noting that $\tr_0(xy)=0$ for $x\in\fh$ and $y\in\fn_\pm$, and that $\fh$ is abelian, this can be simplified to
\[ \tr_0(\partial(h^{-2}\ell^{-1})\ell h^2\wedge {\tilde\partial}uu^{-1}) - 2\tr_0(\partial h h^{-1}\wedge {\tilde\partial}h h^{-1})\,. \]
We can rewrite this, observing that
\[ \tr_0(\partial(h^{-2}\ell^{-1})\ell h^2\wedge{\tilde\partial}(h^{-2}\ell^{-1})\ell h^2) = 4\tr_0(\partial hh^{-1}\wedge{\tilde\partial}hh^{-1})\,, \]
as
\[ \tr_0(\partial(h^{-2}\ell^{-1})\ell h^2\wedge {\tilde\partial}uu^{-1}) - \frac{1}{2}\tr_0(\partial(h^{-2}\ell^{-1})\ell h^2\wedge{\tilde\partial}(h^{-2}\ell^{-1})\ell h^2)\,. \]
We therefore obtain
\[ \frac{1}{\la\alpha_-\,\alpha_+\ra}\int_{\E^4}\mu_{\alpha_-,\alpha_+}\wedge\tr_0\bigg(\partial LL^{-1}\wedge{\tilde\partial} U U^{-1} - \frac{1}{2}\partial LL^{-1}\wedge{\tilde\partial}LL^{-1}\bigg) \]
as required.


\subsection{Novel actions for anti-self-dual Yang-Mills theory} \label{app:SNovel}

In section \ref{sec:sec3} we derive two actions for the ASDYM equations, both appearing in the literature, from HCS on $\PT$. It is natural to speculate that alternative choices of $\Omega$ will lead to novel actions for ASDYM theory. In this appendix we demonstrate that this is indeed the case in two simple examples. Unfortunately the actions we obtain break Lorentz invariance more seriously than those appearing in the bulk of the paper.\footnote{More precisely, the actions appearing in section \ref{sec:sec3} preserve right-handed rotations, whereas those considered in this section preserve neither right-handed nor left-handed rotations. This is an immediate consequence of the boundary conditions on $\bar\cA$, which depend explicitly on choices of right-handed spinors.}\\
	
We begin by considering
\[ \Omega = \frac{\Diff^3Z}{(Z\cdot A)^3(Z\cdot B)} = \frac{\la\pi\diff\pi\ra\wedge\diff^2x^{A'B'}\pi_{A'}\pi_{B'}}{2\la\pi\,\alpha\ra^3\la\pi\beta\ra}\,. \]
and (w.l.o.g.) we assume $\la\alpha\,\beta\ra=1$. To eliminate boundary terms generated when varying the action we must impose boundary conditions at $\pi=\alpha,\beta$. The simplest term to deal with is the one generated at $\pi = \beta$. We have already seen that it is given by
\[ \delta S_\Omega =  -\frac{1}{2}\int_{\E^4}\mu_{\beta,\beta}\wedge\tr(\delta{\bar\cA}\wedge{\bar\cA})|_{\pi={\beta}}\,. \]
As in subsection \ref{subsec:Trig}, we could choose to eliminate this term by imposing trigonometric boundary conditions on ${\bar\cA}$ at $\pi = \beta$. Instead, fixing a right-handed spinor $\nu^A$, we require that both $\nu^A{\bar\cA}_A$ and ${\bar\cA}_0$ are divisible by $\la\pi\,\beta\ra$. We will sometimes refer to a boundary condition of this form at a simple pole as `chiral' by analogy with similar boundary conditions introduced in \cite{costello2019gauge} for \CSt{4}. The boundary term generated at the triple pole is marginally less straightforward to deal with. To understand the contribution of this term it is best to excise a small disk of radius $\epsilon$ around $\pi=\alpha$ on the fibres of $\PT\to\E^4$. Denoting the resulting $\CP^1\setminus D^2_{\alpha,\epsilon}$ bundle over $\E^4$ by $\PT_\epsilon$, we then have
\[ \begin{aligned}
\delta S_\Omega &= \frac{1}{2\pi i}\lim_{\epsilon\to 0}\int_{\PT_\epsilon}\diff(\Omega\wedge \tr(\delta\bar\cA\wedge\bar\cA)) \\
&= \frac{1}{4\pi i}\lim_{\epsilon\to 0}\oint_{S^1_{\alpha,\epsilon}}\frac{\la\diff\pi\,\pi\ra}{\la\pi\,\alpha\ra^3\la\pi\,\beta\ra} \int_{\E^4}\diff^2x^{A'B'}\pi_{A'}\pi_{B'}\wedge\tr(\delta {\bar\cA}\wedge {\bar\cA})\,,
\end{aligned} \]
where $S^1_{\alpha,\epsilon}$ is a circle of radius $\varepsilon$ around $\pi = \alpha$. To eliminate this term we require that $\bar\cA$ be divisible by $\la\pi\,\alpha\ra$, and, choosing another right-handed spinor $\mu^A$, that $\mu^A{\hat\cA}_A$ and ${\bar\cA}_0$ be divisible by $\la\pi\,\alpha\ra^2$. For simplicity we assume that $[\mu\,\nu]=1$.\\
	
Now we fix the gauge. Writing ${\bar\cA}_x = \sigha^{-1}{\bar\partial}_{{\CP}_x}\sigha$,
there are now two pieces of gauge invariant data that can be extracted from ${\bar\cA}_x$. The first is the holomorphic Wilson line from $\beta$ to $\alpha$
\be \cW_{\beta\to\alpha} = {\hat\sigma}^{-1}|_{\pi=\alpha}{\hat\sigma}|_{\pi=\beta} = \sigma\,, \label{eq:3.4Sig} \ee
and the second is the holomorphic derivative of $\sigha$ at $\pi=\alpha$,
\be \la\pi\,\beta\ra^{-2}\sigha^{-1}\p_0\sigha|_{\pi=\alpha} = \chi\,. \label{eq:3.4Chi} \ee
Together these two pieces of data completely characterise $\sigha$ up to the familiar redundancies $\sigha\mapsto h\sigha g^{-1}$. We can therefore fix these ambiguities by choosing a convenient $\sigha$ such that \eqref{eq:3.4Sig} and \eqref{eq:3.4Chi} hold. Writing $\sigha = \sigha_\chi\sigha_\sigma$ we take $\sigha_\sigma$ to be equal to the identity in a neighbourhood, $\cU$, of $\pi=\alpha$ and equal to $\sigma$ in a neighbourhood of $\pi=\beta$. We take $\sigha_\chi$ to be the identity in the complement of $\cU$, so that $\sigha_\chi$ and $\sigha_\sigma$ commute. Inside $\cU$ we choose $\sigha_\chi$ to be\footnote{The rather awkward expression appearing here is the result of using a geometry on spinors in which the points $\alpha$ and $\beta$ are antipodal. This may not be the geometry induced by the Euclidean structure on space-time, but we are only using it to fix the gauge. The effective space-time action will be independent of this choice.}
\[ \sigha_\chi = \exp\left(\frac{\la\pi\,\alpha\ra\overline{\la\pi\,\beta\ra}}{|\la\pi\,\alpha\ra|^2 + |\la\pi\,\beta\ra|^2}f(\pi)\chi\right)\,, \]
where $f$ is a bump function with support inside $\cU$, and equal to $1$ in a neighbourhood of $\alpha$.\\
	
Since $\Omega$ is nowhere vanishing we have
\[ \bar\cA = \sigha^{-1}\bar\p\sigha + \sigha^{-1}\bar\cA'\sigha \]
for $\bar\cA' = {\hat e}^A\pi^{A'}A_{AA'}$, and our boundary conditions fix
\[ A_{AA'} = -\alpha_{A'}(\mu_A\nu^B\beta^{B'}\p_{BB'}\sigma\sigma^{-1} + \nu_A\mu^B\alpha^{B'}\p_{BB'}\chi)\,. \]
We can see that the ASD connection $A$ is in a mixture of the Yang matrix and LMP forms.\\
	
The next step is to substitute our expression for $\bar\cA$ into the action, which gives
\[ \begin{aligned}
S[\sigma,\chi] = \int_{\E^4}\vol_4\,\mu^A\nu^B\tr(\alpha^{A'}\alpha^{B'}\partial_{AA'}\chi\partial_{BB'}\chi - 2\alpha^{A'}\beta^{B'}\partial_{AA'}\chi J_{BB'} \\ + \beta^{A'}\beta^{B'}J_{AA'}J_{BB'}) + \frac{1}{6}\int_{\E^4\times[0,1]}\mu_{\beta,\beta}\wedge\tr\big({\tilde J}^3\big)\,, \end{aligned} \]
where as usual $J = - \diff\sigma\sigma^{-1}$ and ${\tilde J} = - \tilde\diff{\tilde\sigma}{\tilde\sigma}^{-1}$ for $\tilde\sigma$ a smooth homotopy from $\sigma$ to ${\rm id.}$. Consider the associated classical equations of motion. Varying $\chi$ gives
\[ \mu^A\nu^B\alpha^{A'}\p_{AA'}(\alpha^{B'}\p_{BB'}\chi - \beta^{B'}J_{BB'}) = 0\,. \]
Under the assumption that $\chi$ decays and $\sigma$ tends to {\rm id.} at infinity, this equation can be integrated to give
\[ \nu^A(\alpha^{A'}\p_{AA'}\chi - \beta^{A'}J_{BB'}) = 0\,. \]
Varying $\sigma$ gives
\[ \mu^B\nu^A\beta^{A'}\p_{AA'}(\alpha^{B'}\p_{BB'}\chi - \beta^{B'}J_{BB'}) = 0\,, \]
which can be integrated
\[ \mu^A(\alpha^{A'}\p_{AA'}\chi - \beta^{B'}J_{BB'}) = 0\,. \]
Therefore our equations of motion imply that
\[ \alpha^{A'}\p_{AA'}\chi - \beta^{B'}J_{BB'} = 0\,. \]
Applied to $A$ we learn that
\[ \begin{aligned}
A &= \diff x^{AA'}\alpha_{A'}(\mu_A\nu^B\beta^{B'}J_{BB'} - \nu_A\mu^B\alpha^{B'}\partial_{BB'}\chi) \\
&= - \diff x^{AA'}\alpha_{A'}\alpha^{B'}\p_{AB'}\chi = - \diff x^{AA'}\alpha_{A'}\beta^{B'}J_{BB'} \end{aligned} \]
We previously noted that $A$ was in a mixture of the Yang matrix and LMP forms. The equations of motion are essentially the consistency conditions for being able to write $A$ in these two forms.\\
	
In subsection \ref{subsec:Trig} we chose
\[ \Omega = \frac{\Diff^3Z}{\prod_{i=1}^4(A_i\cdot Z)} = \frac{\la\pi\diff\pi\ra\wedge\diff^2x^{A'B'}\pi_{A'}\pi_{B'}}{2\prod_{i=1}^4\la\pi\,\alpha_i\ra}\,, \]
and imposed trigonometric boundary conditions at the simple poles. Here we briefly consider the possibility of introducing chiral boundary conditions instead, {\it i.e.}, fixing a set of right-handed spinors $\{\mu_i\}_{i=1}^4$ we demand that $\mu^A_i{\hat\cA}_A$ and ${\bar\cA}_0$ are divisible by $\la\pi\,\alpha_i\ra$ for $i=1,\dots,4$.\\
	
In the usual way we let ${\bar\cA}_x = \sigha^{-1}{\bar\partial}_{{\CP}_x}\sigha$. The gauge invariant data that can be extracted from ${\bar\cA}_x$ is furnished by the holomorphic Wilson lines between the $\alpha_i$, and can be completely characterised by the map
\[ \sigma:\E^4\to G\backslash G^4\,,\quad x\mapsto [(\sigma_1,\dots,\sigma_4)]\sim [(h\sigma_1,\dots,h\sigma_4)] \]
with $\sigma_i = \hat\sigma|_{\gamma_i}$. We could identify $G\backslash G^4$ with $G^3$ by fixing $\sigma_4={\rm id.}$, but we will find it convenient not to do so. We assume $\sigha$ to be take the value $\sigma_i$ in a neighbourhood of $\alpha_i$ for all $i=1,\dots,4$. Then the gauge invariant holomorphic Wilson lines are
\[ \cW_{\gamma_i\to\gamma_j} = \sigha^{-1}|_{\gamma_j}\sigha|_{\gamma_i} = \sigma_j^{-1}\sigma_i\,. \]
Since $\Omega$ is free from zeros
\[ \bar\cA = \sigha^{-1}\bar\p\sigha + \sigha^{-1}\bar\cA'\sigha \]
for $\bar\cA' = {\hat e}^A\pi^{A'}A_{AA'}$. Our boundary conditions then imply that
\[ A_a = \sum_{i=1}^4 u^{(i)}_a\mu_i^B\alpha_i^{B'}{J_i}_{BB'}\,, \]
where $J_i = -\diff\sigma_i\sigma_i^{-1}$ and $u^{(i)}_a$ is the dual basis to $\mu_i^A\alpha_i^{A'}$, i.e. $u^{(i)}_{AA'}\mu_j^A\alpha_j^{A'} = \delta^i_{\,j}$. Substituting this expression for $\bar\cA$ back into the action gives
\[ \begin{aligned}
\sum_{i,j=1}^4\int_{\E^4}\vol_4\,g^{AB}_{ij}\alpha_i^{A'}\alpha_j^{B'}\tr({J_i}_{AA'}{J_j}_{BB'}) + \sum_{i=1}^4\frac{h_i}{3}\int_{\E\times[0,1]}\mu_{\alpha_i,\alpha_i}\wedge\tr({\tilde J}_i^3)\,.
\end{aligned} \]
As usual $J_i = -\diff\sigma_i\sigma_i^{-1}$, ${\tilde J}^\alpha = -{\tilde\diff}{\tilde\sigma}_i{\tilde\sigma}_i^{-1}$ for $\tilde\sigma_i$ a smooth homotopy from $\sigma_i$ to ${\rm id.}$. The couplings $g^{AB}_{ij} = g^{BA}_{ji}$ and $h_i$ are given by
\[ h_i = \prod_{j=1,\,j\neq i}^4\frac{1}{\la\alpha_i\,\alpha_j\ra}\,,\quad g^{AB}_{ii} = \frac{h_i}{\Delta}\mu_i^{(A}\nu_i^{B)}\,,\quad g_{ij}^{AB} = \frac{[\mu_{i'}\,\mu_{j'}]}{\la\alpha_i\,\alpha_j\ra\Delta}\mu_i^A\mu_j^B~{\rm for}~ i\neq j\,, \]
Here $i'$ and $j'$ are determined by the requirement that $(i,j,i',j')$ is an even permutation of $(1,2,3,4)$, and we have introduced
\[ \begin{aligned} \Delta &= [\mu_1\,\mu_2][\mu_3\,\mu_4]\la\alpha_1\,\alpha_4\ra\la\alpha_2\,\alpha_3\ra - [\mu_1\,\mu_4][\mu_2\,\mu_3]\la\alpha_1\,\alpha_2\ra\la\alpha_3\,\alpha_4\ra\,, \\
\nu_1^A &= \mu_4^A[\mu_2\,\mu_3]\la\alpha_1\,\alpha_2\ra\la\alpha_3\,\alpha_4\ra - \mu_2^A[\mu_3\,\mu_4]\la\alpha_1\,\alpha_4\ra\la\alpha_2\,\alpha_3\ra\,,
\end{aligned}\]
with the remaining $\nu_i$ obtained via cyclic permutation weighted by sign. Note that under the natural action of $S_4$ on $\Delta$ it transforms in the sign representation. Similarly $\nu_i^A$ transforms in the sign representation of $S^3$.\\
	
Under symmetry reduction by a 2d group of translations both actions appearing in this section will give theories with equations of motion equivalent to the principal chiral model with WZW term. We will not perform these reductions explicitly here. These same reductions can also be performed on twistor space and will give rise to \CSt{4} with appropriate insertions of disorder defects. It is notable that such a large class of disorder defects give actions for the same integrable system.


\vspace{1cm}

\noindent {\large{\bf Acknowledgments}}

It is a pleasure to thank Tim Adamo, Kevin Costello and Beno\^{i}t Vicedo for helpful discussions. This work has been partially supported by STFC consolidated grants ST/P000681/1, ST/T000694/1. The work of RB is supported by EPSRC studentship EP/N509620/1.

\vspace{1cm}

\bibliographystyle{JHEP}
\bibliography{references}

\end{document}